\documentclass[useAMS,usenatbib,a4paper]{mn2e} 
\usepackage{amsmath,graphics,color,subfigure,epsfig,amssymb}
\usepackage{mathptmx}
\usepackage{epsfig}
\usepackage{booktabs}
\usepackage{rotating,rotfloat}
\usepackage{longtable}
\usepackage{graphicx}
\usepackage{url} 
\usepackage{listings}
\usepackage{color}
\usepackage{pslatex} 
\usepackage{booktabs}
\usepackage{dcolumn}
\usepackage{widetext}
\usepackage[hyperindex,breaklinks=true, colorlinks, citecolor=blue]{hyperref}

\newcommand{\adsurl}[1]{\href{#1}{ADS}} 
\providecommand{\url}[1]{\href{#1}{#1}}

\let\la=\lesssim    
\def\reff@jnl#1{{\rm#1\/}}
\def\aj{\reff@jnl{AJ}}                  
\def\araa{\reff@jnl{ARA\&A}}            
\def\actaa{\reff@jnl{Acta. Astron}}     
\def\apj{\reff@jnl{ApJ}}                
\def\apjl{\reff@jnl{ApJ}}               
\def\apjs{\reff@jnl{ApJS}}              
\def\ao{\reff@jnl{Appl.Optics}}         
\def\apss{\reff@jnl{Ap\&SS}}            
\def\aap{\reff@jnl{A\&A}}               
\def\aapr{\reff@jnl{A\&A~Rev.}}         
\def\aaps{\reff@jnl{A\&AS}}             
\def\azh{\reff@jnl{AZh}}                
\def\baas{\reff@jnl{BAAS}}              
\def\jrasc{\reff@jnl{JRASC}}            
\def\memras{\reff@jnl{MmRAS}}           
\def\mnras{\reff@jnl{MNRAS}}            
\def\pra{\reff@jnl{Phys.Rev.A}}         
\def\prb{\reff@jnl{Phys.Rev.B}}         
\def\prc{\reff@jnl{Phys.Rev.C}}         
\def\prd{\reff@jnl{Phys.Rev.D}}         
\def\prl{\reff@jnl{Phys.Rev.Lett}}      
\def\pasp{\reff@jnl{PASP}}              
\def\pasj{\reff@jnl{PASJ}}              
\def\qjras{\reff@jnl{QJRAS}}            
\def\skytel{\reff@jnl{S\&T}}            
\def\solphys{\reff@jnl{Solar~Phys.}}    
\def\sovast{\reff@jnl{Soviet~Ast.}}     
\def\ssr{\reff@jnl{Space~Sci.Rev.}}     
\def\zap{\reff@jnl{ZAp}}                
\def\nat{\reff@jnl{Nature}}             
\def\pasa{\reff@jnl{Publ. Astron. Soc. Aust.}}            
\def\jcap{\reff@jnl{Journal of Cosmology and Astroparticle Physics}}     
\def\jqsrt{\reff@jnl{Journal of Quantitative Spectroscopy \& Radiative Transfer}}
\def\nar{\reff@jnl{New Astronomy Reviews}}

\newcolumntype{K}{D{.}{.}{2,2}}
\newcolumntype{H}{D{,}{\pm}{4.4}}

\newcommand{\wmap}{{\em WMAP~}}
\newcommand{\planck}{{\em Planck~}}

\newcommand{\dg}{\mbox{$^{\circ}$}}
\newcommand{\beq}{\begin{equation}}
\newcommand{\eeq}{\end{equation}}
\newcommand{\vd}{{\bdelta}}

\newcommand{\kband}{K--band}
\newcommand{\kaband}{Ka--band}
\newcommand{\qband}{Q--band}

\newcommand{\nside}{N_{\text{side}}} 

\newcommand{\pnaive}{\hat p'}
\newcommand{\pka}{\hat p_{\chi}}
\newcommand{\pwk}{\hat p_{\rm AS}}
\newcommand{\pmas}{\hat p_{\rm MAS}}

\title[A new polarisation amplitude bias reduction method.]{A new
  polarisation amplitude bias reduction method.}

\author[Vidal, Leahy \& Dickinson]{Matias~Vidal$^{1,2}$\thanks{E-mail:
    mvidal@das.uchile.cl}, J.~P.~Leahy$^{1}$\thanks{E-mail:
    j.p.leahy@manchester.ac.uk} and C.~Dickinson$^{1}$\thanks{E-mail:
    clive.dickinson@manchester.ac.uk} \\ $^{1}$Jodrell Bank Centre for
  Astrophysics, Alan Turing Building, School of Physics and Astronomy,
  The University of Manchester,\\ Oxford Road, Manchester M13 9PL,
  U.K. \\$^{2}$Departamento de Astronom\'{i}a, Universidad de Chile, Casilla 36-D, Santiago, Chile. }

\begin{document}


\pagerange{\pageref{firstpage}--\pageref{lastpage}} \pubyear{2015}

\maketitle

\label{firstpage}

\begin{abstract}

Polarisation amplitude estimation is affected by a positive noise
bias, particularly important in regions with low signal-to-noise ratio
(SNR). We present a new approach to correct for this bias in the case
there is additional information about the polarisation angle. We
develop the `known-angle estimator' that works in the special case
when there is an independent and high signal-to-noise ratio ($\gtrsim
2\sigma$) measurement of the polarisation angle. It is derived for the
general case where the uncertainties in the $Q,U$ Stokes parameters
are not symmetric. This estimator completely corrects for the
polarisation bias if the polarisation angle is perfectly known. In the
realistic case, where the angle template has uncertainties, a small
residual bias remains, but that is shown to be much smaller that the
one left by other classical estimators. We also test our method with
more realistic data, using the noise properties of the three lower
frequency maps of \wmap\!\!. In this case, the known-angle estimator
also produces better results than methods that do not include the
angle information. This estimator is therefore useful in the case
where the polarisation angle is expected to be constant over different
data sets with different SNR.

\end{abstract}
\begin{keywords}
polarisation -- techniques: polarimetric -- methods: data analysis --
methods: statistical -- radiation mechanisms: non-thermal
\end{keywords}

\section{Introduction}

It has long been noticed that observations of linear polarisation are
subject to a positive bias \citep{serkowski:58}. Given the positive
nature of the polarisation amplitude, $P=\sqrt{Q^2+U^2}$, even if the
true Stokes parameters $Q_0, U_0$ are zero, $P$ will yield a non-zero
estimate in the presence of noise. The effect is particularly
important in the low signal-to-noise ratio (SNR) regime.  Ways to
correct for the bias have been studied in detail for the special case
where the uncertainties for $(Q,U)$ are equal and normally distributed
around their true value $(Q_0,U_0)$
\citep{wardle:74,simmons:85,vaillancourt:06,quinn:12}.
\citet{montier:14a, montier:14b} give a useful review comparing
different bias reduction methods.

In this paper we propose a new approach, useful when there is an
independent measurement of the polarisation angle
$\chi=0.5\arctan(U/Q)$. This situation occurs, for instance, in the
polarisation datasets from \wmap \citep{bennett:13} and \planck
\citep{planck_results2013}, where over a range of frequency the
polarisation angle is expected to be nearly constant (e.g.
synchrotron radiation will have the same polarisation angle at
different frequencies as this angle depends on the direction of the
magnetic field of the emitting medium), and the variation of the
polarised intensity with frequency is of interest.  Using simulations,
we test the performance of this estimator compared to previous methods
from the literature that do not include angle information.  In
Section~\ref{sec:methods} we show the origin of the bias and we
describe some methods used to correct for it. In
Section~\ref{sec:known_angle} we derive our new estimator, its
uncertainty and residual bias. In Section~\ref{sec:tests} we use
simulations to test its performance. Section~\ref{sec:conclusions}
concludes.


\section{Bias correction methods}
\label{sec:methods}
We will first review the simple case when the uncertainties in $Q$ and
$U$ are equal, and then study the more general case with asymmetric
uncertainties.

\subsection{Symmetric uncertainties}

Let us take $(Q_0,U_0)$ as the true Stokes parameters from a source
and $(Q',U')$ the measured ones. We can write the joint probability
distribution function (p.d.f.) for $(Q',U')$ as the product of the
individual normal distributions
\begin{align}
  f(Q',U') &= \frac{1}{2 \pi \sigma^2} \exp\left[ -
  \frac{(Q'-Q_0)^2+(U'-U_0)^2}{2\sigma^2} \right],
  \label{eq:joint_no_corr}
\end{align}
where $\sigma_Q=\sigma_U=\sigma$ is the uncertainty in $Q'$ and $U'$.

Transforming into polar coordinates using the definitions for the
polarisation amplitude $P=\sqrt{Q^2+U^2}$ and the polarisation angle $\chi=\frac{1}{2}\arctan{(U/Q)}$, we have:
\begin{align}
  f(P',\chi') =& \frac{P'}{\pi \sigma^2} 
  e^{ - [P'^2 + P_0^2 - 2(P'\cos2\chi' P_0\cos2\chi_0 + P\sin2\chi' P_0\sin2\chi_0 )]/ 2\sigma^2} \nonumber \\
  =& \frac{P'}{\pi \sigma^2}
  e^{- (P'^2 + P_0^2) /2\sigma^2}
  e^{-P'P_0\cos[2(\chi' -\chi_0)]/\sigma^2}.
  \label{eq:joint_no_corr_pol} 
\end{align}
The marginal probability distribution for $P'$ is obtained by
integrating $f(P',\chi')$ over $\chi'$. This angular integral can be
written as a function of the modified Bessel function of first type
$I_0(z)$, yielding the Rice distribution for polarisation
\citep{Rice1945}:
\begin{equation}
  R(P'|P_0)=\frac{P'}{\sigma^2}I_0\left(\frac{P'P_0}{\sigma^2}\right)e^{-(P'^2+P_0^2)/2\sigma^2}.
  \label{eq:rice}
\end{equation} 

\begin{figure}
  \centering
  \includegraphics[angle=0,width=0.48\textwidth]{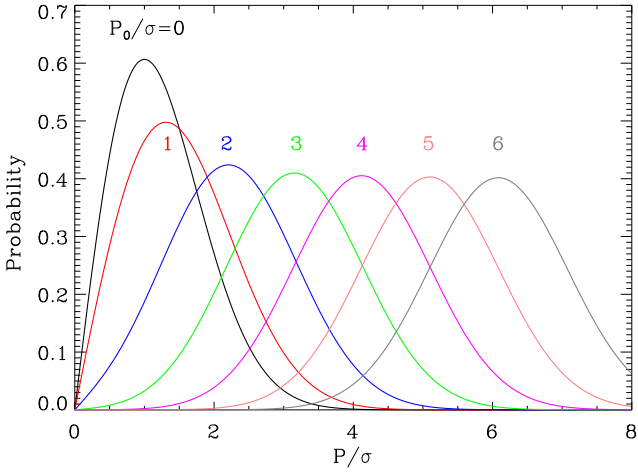}
  \caption[Rice distribution for different SNR values.]{Rice
    distribution (Eq. \ref{eq:rice}) plotted for different values of
    the true SNR, $P_0/\sigma$. The asymmetry and bias are clear in
    the low SNR level. At high SNR, the distribution converges to a
    Gaussian with standard deviation $\sigma$ centred at
    $\sqrt{P_0+\sigma^2}$.}
  \label{fig:rice_dist}
\end{figure}

It is important to note that the integral of $R(P'|P_0)$ represents the
probability of measuring $P'$ inside an interval for a given true
polarisation $P_0$. Fig. \ref{fig:rice_dist} shows $R(P'|P_0)$ for
different SNR. The bias, defined as $\langle P'\rangle-P_0$, arises because
this probability distribution is not symmetric, and becomes clear in
Fig.~\ref{fig:rice_dist} at low SNR. Even when the true SNR is
zero (black curve in the figure), the measured value is close to 1. At
large SNR, the distribution approaches a Gaussian with
mean close to $P_0$ and standard deviation close to $\sigma$.

\citet{simmons:85} compared five estimators for $P_0$ (including the
uncorrected $P'$), and concluded that the one with smallest residual
bias when $P_0/\sigma \gtrsim 0.7$ is that suggested by
\citet{wardle:74}:
\begin{align}
  \frac{\partial R}{\partial P'}(P',P_0)\bigg|_{P_0=\hat{p}} = 0.
  \label{eq:cond_wk}
\end{align}
(note that this is {\em not} the maximum likelihood estimator). Even
at moderate SNR, the rice distribution is not centred at the measured
SNR, this can be seen in Fig.  \ref{fig:rice_dist}, where the
distributions have mean $\sqrt{P_0+\sigma^2}$. This motivated
\citet{wardle:74} to propose a very simple estimator
\begin{equation}
  \pwk =\left\{ \begin{array}{lr} \sqrt{P'^2-\sigma^2} & P'
    \ge \sigma; \\ 0 & \mbox{otherwise}.
  \end{array} \right.
  \label{eq:p_wk}
\end{equation}
This approximates Eq.\,\ref{eq:cond_wk}, and also has the virtue of
giving lower bias at very low SNR, albeit at the cost of a 1 per cent
over-correction near SNR = 2; this has been widely used in
practice. \citet{montier:14b} compared various estimators of the
polarisation angle in terms of the residual bias, risk, variance and
Gaussianity.  An unwanted property of the asymptotic (AS,
Eq. \ref{eq:p_wk}) estimator is that the distribution of the estimator
is discontinuous as it yields zeroes below a particular SNR (see
Eq. \ref{eq:p_wk}). It is illuminating to re-write Eq.\,\ref{eq:p_wk}
in terms of the error in the polarisation angle, $\sigma_\chi =
\sigma/2P'$, so
  \begin{align}
    \pwk =  P' \sqrt{1-4\sigma_\chi^2}.
    \label{eq:pwk_chi}
  \end{align}
This emphasises that the source of the bias is the error in the angle of the
$(Q,U)$ vector, which contributes a component of the error
vector orthogonal to the true polarisation with length
$P'\sin2\Delta_\chi$, and is added in quadrature to the parallel component,
hence always contributing a positive bias.

For the polarisation angle, here we use the naive estimator, $\chi =
\frac{1}{2}\arctan(U/Q)$, which is completely unbiased. The
uncertainty of this estimator, on the asymptotic case where
$P_0\gg\sigma$ is \citep{vinokur:65,montier:14a}
\begin{equation}
  \sigma_{\chi}=\sqrt{Q'^2\sigma^2_Q +U'^2\sigma^2_U}/2P^2.
  \label{eq:sigmachigauss}
\end{equation}


\subsection{Asymmetric uncertainties}
\label{sec:pol_bias_chi}

\begin{figure}
  \centering
  \includegraphics[angle=0,width=0.48\textwidth]{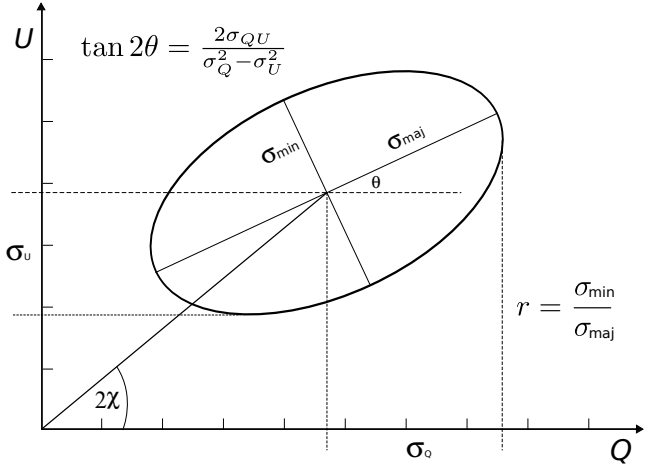}
  \caption{Error ellipse in the $(Q,U)$ plane showing how the
    polarisation angle $\chi$ and the rotation angle $\theta$ are
    related.}
  \label{fig:error_ellipse}
\end{figure}

The previous case in which the uncertainties in $(Q',U')$
are equal and uncorrelated is well understood. The
asymmetric case is interesting as many
polarisation data sets have this characteristic. For example, the CMB
experiments \wmap \citep{bennett:13} and \planck
\citep{planck_results2013} have correlations between the $(Q',U')$
uncertainties due to non-uniform azimuthal coverage for each pixel in
the sky. This case has been recently studied by 
\citet{montier:14a,montier:14b,plaszczynski:14}.
The error distribution in these cases is an elliptical 2D
Gaussian in $(Q',U')$, characterised by a covariance matrix \citep[e.g.][]{kendall:1994}
\begin{equation}
 C = \left(\begin{array}{cc} \sigma_Q^2 & \rm{cov_{QU}} \\ 
                             \rm{cov_{QU}} & \sigma_U^2 \end{array}\right)
\end{equation}
Defining a $(Q,U)$ error vector
\begin{equation}
\vd = \left( \begin{array}{c} Q' - Q_0 \\ U'-U_0 \end{array}\right),
\end{equation}
we have
\begin{align}
  f(Q',U') =&  {1 \over 2\pi \sqrt{\det[C]}} 
                \exp\left[-{\vd^T C^{-1} \vd \over 2} \right]~~~~~~~~~~~~~~~~~~~~~~~~~~~~~~~~~~~~~~~~\nonumber \\
 =& \frac{1}{2 \pi  \sigma_Q \sigma_U \sqrt{1-\rho^2}} \exp\biggl[ -\frac{1}{2(1-\rho^2)}\biggl( \nonumber \\
     & \frac{(Q'-Q_0)^2}{\sigma_Q^2} +
      \frac{(U'-U_0)^2}{\sigma_U^2} -
      \frac{2\rho(Q'-Q_0)(U'-U_0)}{\sigma_Q \sigma_U}
      \biggr)\biggr],
  \label{eq:pdf_qu}
\end{align}

where $\rho$ is the correlation coefficient between $(Q_0,U_0)$,
\begin{align}
  \rho &=\frac{E[(Q'-Q_0)(U'-U_0)]}{\sigma_Q\sigma_U} =
  \frac{\rm{cov_{QU}}}{\sigma_Q\sigma_U}.\nonumber \\
\end{align}
The error ellipse in the $Q,U$ plane will be rotated an angle $\theta$
in the case of a non-zero $\rm{cov_{QU}}$ (see Fig.~\ref{fig:error_ellipse}).
In terms of the components of the covariance matrix,
\begin{equation}
\theta = \frac{1}{2}\arctan{2\rm{cov_{QU}} \over \sigma_Q^2 -
  \sigma_U^2}.
\end{equation}
The error ellipse has axial ratio
\begin{equation}
  r = \frac{\sigma_{\rm min}}{\sigma_{\rm maj}},
\label{eq:r}
\end{equation}
where 

\begin{align}
  \sigma^2_{\rm maj} &= \frac{1}{2} \left( \sigma_Q^2 + \sigma_U^2 +
 \sqrt{(\sigma_Q^2-\sigma_U^2)^2 + 4\rm{cov_{QU}}^2}\,\right) \\
  \sigma^2_{\rm min} &=  \frac{1}{2} \left( \sigma_Q^2 + \sigma_U^2 -
 \sqrt{(\sigma_Q^2-\sigma_U^2)^2 + 4\rm{cov_{QU}}^2}\,\right).
\end{align}

Allowing for correlated variables, the
error in any $f(Q',U')$ is to first order
\begin{equation}
\sigma_f^2 = \left( \partial f \over \partial Q'\right)^2 \sigma_Q^2 + 
2 {\partial f \over \partial Q'}{\partial f \over \partial U'} \rm{cov_{QU}}
+ \left( \partial f \over \partial U'\right)^2 \sigma_U^2.
\label{eq:error_prop}
\end{equation}
Hence \citep{montier:14a},
\begin{eqnarray}
\sigma_P^2 &=& {Q'^2\sigma_Q^2 + 2Q'U'\rm{cov_{QU}} + U'^2\sigma_U^2 \over P'^2} 
\nonumber \\
&=&  \cos^22\chi'\, \sigma_Q^2 + 2\cos2\chi'\sin2\chi'\, \rm{cov_{QU}} + \sin^22\chi' \,\sigma_U^2 \label{eq:sigma_P}\\
\sigma_\chi^2 &=& 
{ \sin^22\chi'\, \sigma_Q^2 - 2\cos2\chi'\sin2\chi'\, \rm{cov_{QU}} + 
\cos^22\chi' \,\sigma_U^2 \over 4 P'^2}, \label{eq:sigma_chi}
\end{eqnarray}

When $r \ne 1$, the polarisation bias depends on the polarisation
angle $\chi_0$, more specifically on $\theta - 2\chi_0$, as well as on
$p_0$. The asymptotic estimator from Eq.\,\ref{eq:pwk_chi}, $\pwk =
P'\sqrt{1-4\sigma_\chi^2}$, still applies, provided we use the
generalised form of $\sigma_\chi$ derived above.  We note that
Eq.\,\ref{eq:sigma_chi} is derived in the small-error approximation
and uses the observed polarisation angle, $\chi'$ as a surrogate for
the true polarisation angle $\chi_0$, which is unobjectionable in the
high SNR regime. When SNR is low it is not so obvious that this is
appropriate, but we will verify by simulations that the estimator
remains reasonably effective.

The preferred estimator of \citet{montier:14b}, is the `modified asymptotic
estimator', MAS  \citep{plaszczynski:14}, which using the above notation is given by:
\begin{equation}
\pmas = P'\left[1 - 2\sigma_\chi^2\left(1-e^{-P'^2/b^2}\right)\right],
\label{eq:p_mas}
\end{equation}


\section{Known-angle estimator}
\label{sec:known_angle}
As we will see, all estimators based exclusively on one observed
$(Q',U')$ measurement give a significant positive bias at very low
SNR. However, this can be overcome if we have an independent estimate
of the polarisation angle $\chi$. We will refer to this as the
`template' observation, in contrast to the `target' observation to be
debiased.

We will build a new estimator, based on the maximum likelihood
estimator, assuming that the true angle $\chi_0$ is know.  We will
later calculate the residual bias which emerges because of uncertainty
in our template angle, hereafter denoted $\chi$, as distinct from the
true angle $\chi_0$ and the observed angle $\chi'$ from the target
observation.

To find our known-angle estimator, $\pka$, we take the joint
p.d.f. for the observed values $(Q',U')$ with asymmetric uncertainties
from Eq. \ref{eq:pdf_qu}. Since $Q_0=P_0\cos2\chi_0$ and
$U_0=P_0\sin2\chi_0$,
\begin{align}
  f(Q',U') =& \frac{1}{2 \pi \sigma_Q \sigma_U \sqrt{1-\rho^2}}   \exp\biggl[ -\frac{1}{2(1-\rho^2)}\biggl(  \nonumber \\ 
    & \frac{(Q' - P_0\cos 2\chi_0 )^2}{\sigma_Q^2} +  \frac{(U' - P_0\sin 2\chi_0)^2}{\sigma_U^2} \nonumber \\ 
    & - \frac{2\rho(Q' - P_0\cos 2\chi_0
      )(U' - P_0\sin 2\chi_0 )}{\sigma_Q \sigma_U} \biggr)\biggr].
\end{align}

The maximum likelihood estimator in this case, $\pka$, is
defined by the condition:
\begin{equation}
  \frac{\partial f(Q',U')}{\partial P_0}\bigg|_{P_0=\pka}=0.
\end{equation}
With the assumption that $\chi_0 = \chi$, 
this leads to the expression for the de-biased polarisation
amplitude,
\begin{align}
    \pka = 
    \frac{
      \sigma^2_U Q'\cos 2\chi
      - \rm{cov_{QU}}(Q'\sin 2\chi +U'\cos{2\chi})
      + \sigma^2_Q U' \sin 2\chi
    }
         {
           \sigma^2_U\cos^22\chi 
           - 2\rm{cov_{QU}}\sin 2\chi \cos 2\chi
           + \sigma^2_Q\sin^22\chi
         }.
  \label{eq:pdeb_chi}
\end{align}

We note that if the observed
polarisation angle $\chi'$ is used as a surrogate for $\chi_0$, then 
$\pka = p' = \sqrt{Q'^2+U'^2}$ and there is no correction
whatsoever -- we do need an independent constraint on $\chi$ to benefit
from this approach.

The error in $\pka$ is given via Eq.\,\ref{eq:error_prop}:
\begin{eqnarray}
    \sigma^2_{\pka}&=&\frac{
      \sigma^2_Q\sigma^2_U - \sigma^2_{QU} } {\sigma^2_U\cos^22\chi - 2\rm{cov_{QU}}\sin 2\chi \cos 2\chi + \sigma^2_Q\sin^22\chi  } \nonumber \\
 &=& { \det[C] \over 4P'^2 \,\sigma_\chi^2}. 
\label{eq:error_pchi}
\end{eqnarray}

The known-angle estimator still contains a residual bias, due to the
uncertainty in the template angle $\chi$. Specifically the expected
value of the fractional bias is
\begin{equation}
b_{\chi} \equiv {\langle \pka \rangle - P_0 \over P_0} = 
\int_\chi f(\chi|\chi_0) \int_{Q',U'} f(Q',U'|P_0,\chi_0)\,  {\pka \over P_0} \,
d\chi\,dQ\,dU - 1.
\end{equation}
Since $\pka$ is linear in $Q'= P_0\cos2\chi_0 + \delta_Q$ and in 
$U' = P_0\sin2\chi_0 + \delta_U$,
integrating over the Gaussian probability distribution of $(Q',U')$ eliminates
the dependence on the the deviations $\delta_Q$, $\delta_U$, and we have

\begin{widetext}
  \begin{equation}
    b_{\chi} = 
    \int_\chi f(\chi|\chi_0) {\sigma_U^2 P_0 \cos2\chi_0 \cos2\chi 
      - \rm{cov_{QU}}P_0(\cos2\chi_0\sin2\chi + \sin2\chi_0\cos2\chi) 
      + \sigma_Q^2 P_0 \sin2\chi_0\sin2\chi \over P_0(\sigma_U^2\cos^22\chi 
      - 2\rm{cov_{QU}} \cos2\chi\sin2\chi + \sigma_Q^2\sin^22\chi)} \, d\chi - 1,
\label{eq:bias_chi}
  \end{equation}
\end{widetext}
where $f(\chi|\chi_0)$ is the p.d.f. of the template angle.

The amplitude of the polarised signal, $P_0$, 
cancels, so the residual bias just depends on the axial ratio of the
$(Q,U)$ error ellipse, $r$, and the difference between the orientation
of the ellipse, $\theta$,
and that of the true $(Q_0,U_0)$ vector, $2\chi_0$ (see Fig.~\ref{fig:error_ellipse}). If the angle is known exactly so $f(\chi|\chi_0) = \delta(\chi-\chi_0)$,
the residual bias vanishes. 

\begin{figure}
  \centering
  \includegraphics[angle=0,width=0.48\textwidth]{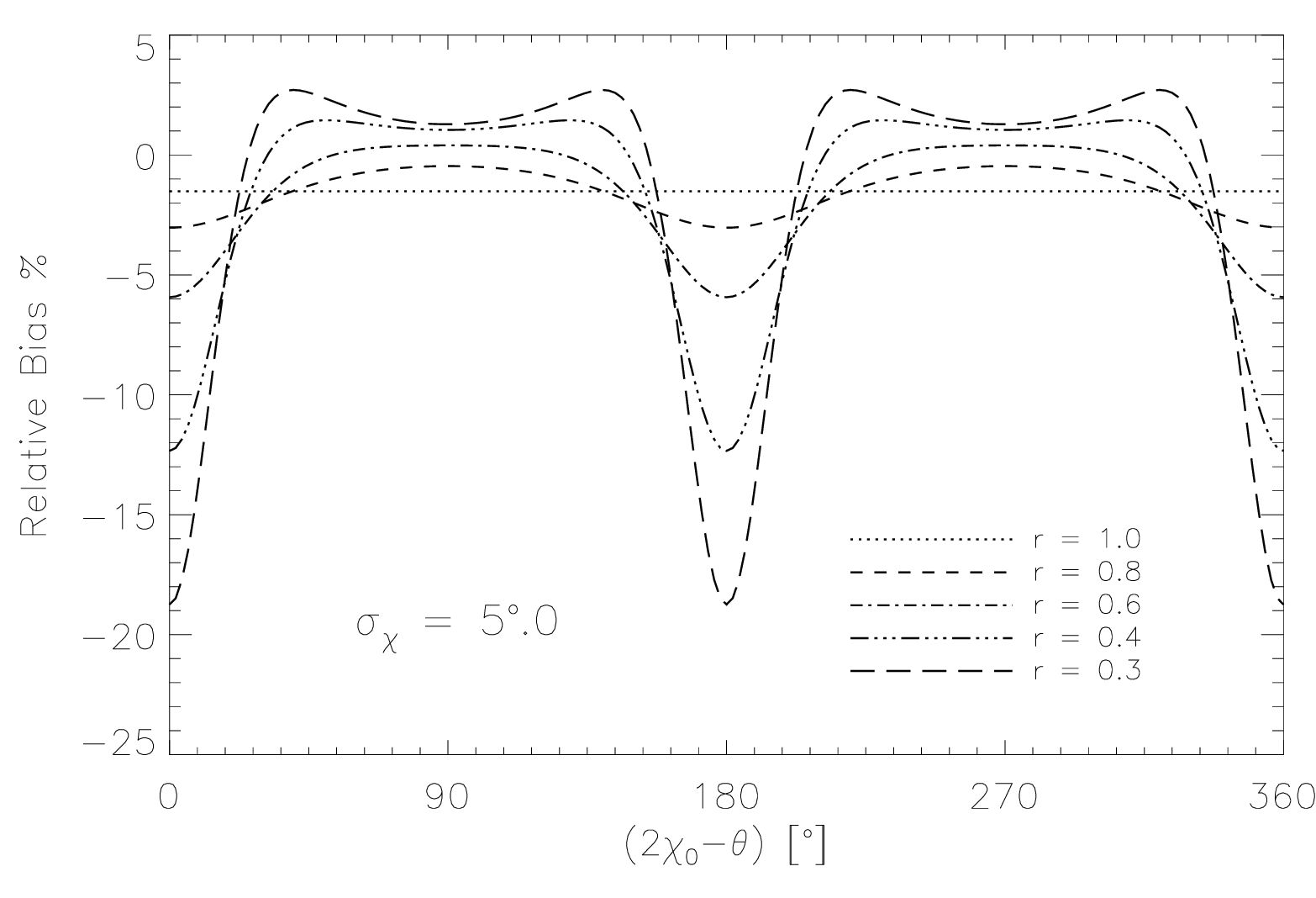}
  \includegraphics[angle=0,width=0.48\textwidth]{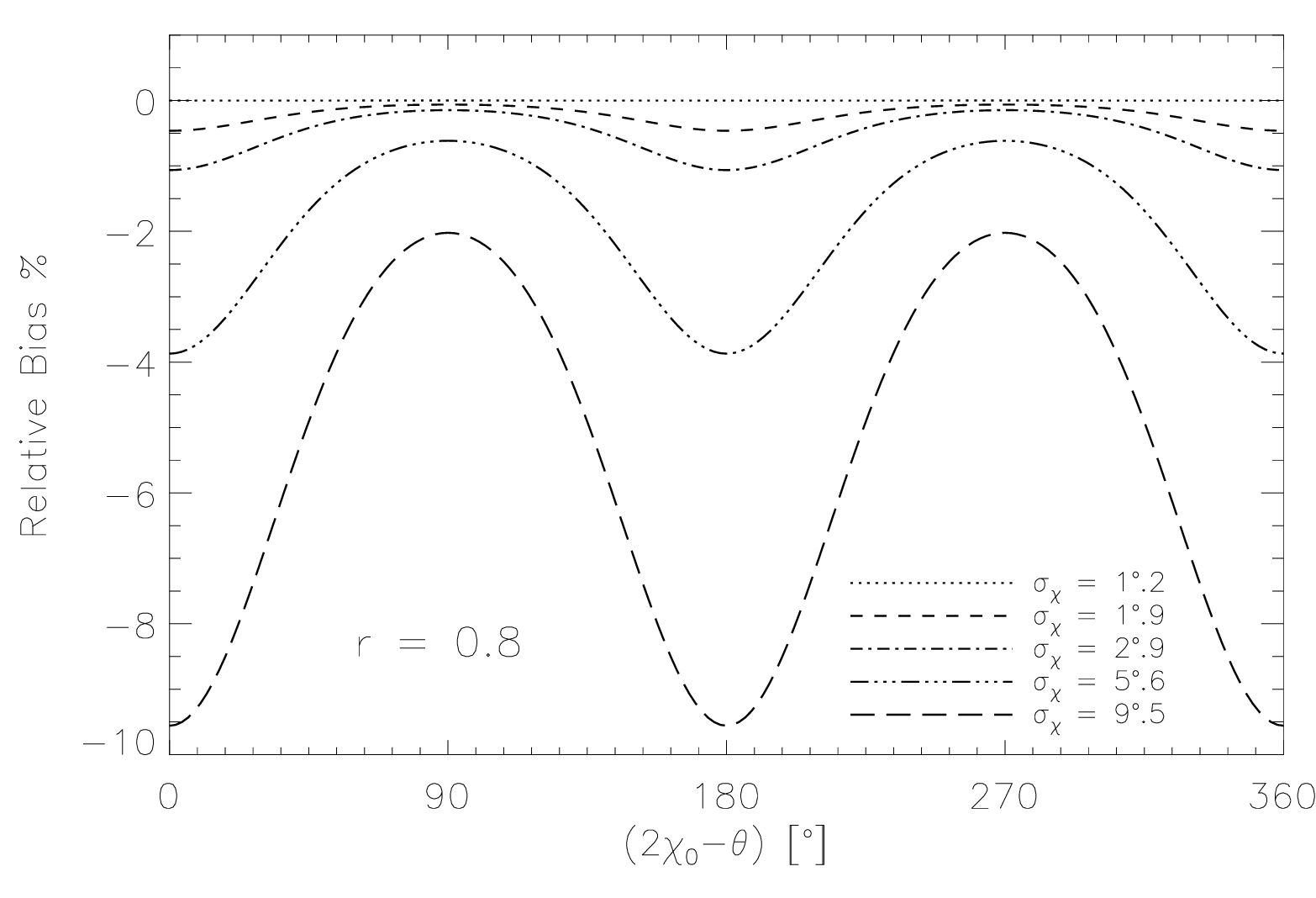}
  \caption{Fractional bias of the known-angle estimator,
    $\pka$, for different values of the error ellipse
    ($r=\sigma_U^{\prime}/\sigma_Q^{\prime}$) and an uncertainty in
    the template angle of 5\dg\ ({\it top}). The {\it bottom} plot shows
    the residual fractional bias of $\pka$ for different values of the
    uncertainty in the template angle, $\sigma_{\chi}$, for
    a fixed value of $r=0.8$. Both plots are calculated with $P_0=1$.}
  \label{fig:residual_bias}
\end{figure}

In the special case where $\rm{cov_{QU}} = 0$, so $r =\sigma_U/\sigma_Q$, we have
\begin{equation}
  b_{\chi} = 
  \int_\chi f(\chi|\chi_0) {r^2\cos2\chi_0 \cos2\chi 
    + \sin2\chi_0\sin2\chi \over r^2\cos^22\chi 
    + \sin^22\chi} \, d\chi - 1.
\label{eq:bias_nocorr}
\end{equation}
The top panel of Fig.~\ref{fig:residual_bias} shows the  bias
for different axial ratios of the error ellipse as function of the
polarisation angle (see the angle definitions in
Fig.~\ref{fig:error_ellipse}).  Here the uncertainty in the template
angle is Gaussian (c.f. Eq. \ref{eq:sigmachigauss}) and it is fixed at $\sigma_\chi = 5$\dg.  If the
error distribution is symmetric ($r = 1$), the  bias is
constant (cf. Eq. \ref{eq:bias_nocorr}).
The bottom panel of Fig.~\ref{fig:residual_bias} shows the bias for
different values of the uncertainty in the template angle, for an
error ellipse with axial ratio $r = 0.5$. The bias is usually
negative, i.e. the polarised intensity is slightly underestimated.

When the template for the angle has an SNR at least 2 times larger than
the target, the known-angle estimator thus gives excellent performance
at very low signal-to-noise levels, where other estimators have large
residual biases.  However, for a given uncertainty in $\chi$, the bias
does not decrease as the signal in $(Q',U')$ rises, unlike all the
standard estimators.
In practical use, the estimator is only worthwhile when a template observation
with substantially higher SNR than the target observations is available;
fortunately, this situation is fairly common, as discussed in 
Section~\ref{sec:wmap_sims}.

\section{Tests of the estimators}
\label{sec:tests}

Here we compare the effectiveness of three bias reduction methods
using Monte Carlo simulations for a range of SNR. This is to show what
can be gained when including additional angle information from a
higher signal-to-noise template. In section~\ref{sec:general} we study
the residual bias in a single pixel for a range of SNR in $(Q,U)$.
Section~\ref{sec:wmap_sims} tests the methods using real noise values
from the \wmap polarisation data.

\begin{figure}
  \centering
  \includegraphics[angle=0,width=0.48\textwidth]{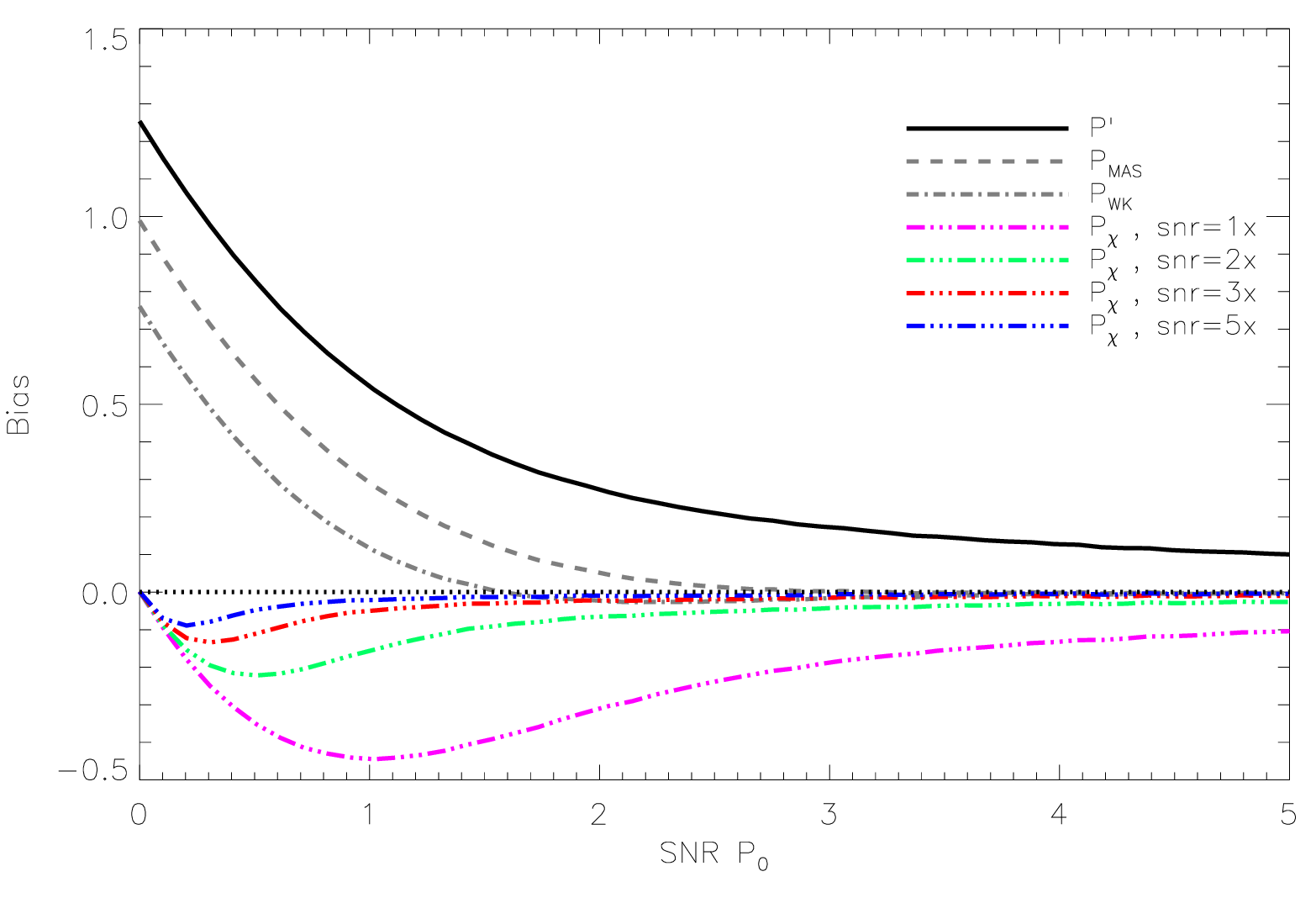}
  \includegraphics[angle=0,width=0.48\textwidth]{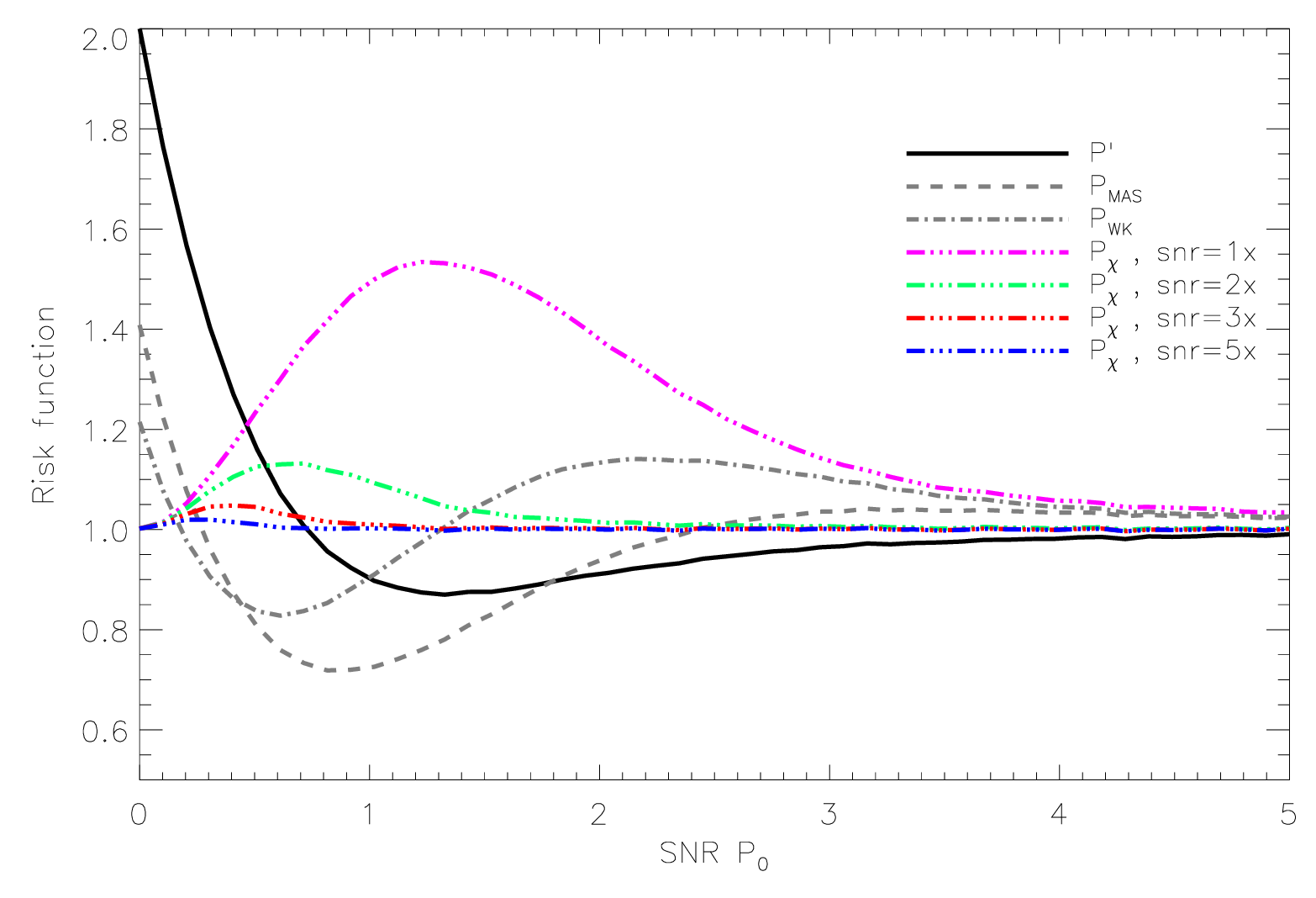}
  \caption{Residual bias ({\it top}) and risk function ({\it bottom})
    for the naive, $\pnaive$ (continuum line), $\pmas$ (dashed),
    $\pwk$ (dot-dashed line) and $\pka$ (colours) estimators as a
    function of the SNR of the true polarisation amplitude $P_0$. As
    the residual bias and risk function of the known-angle estimator
    depends on the uncertainty of the high SNR template, we show these
    two functions for four different values of the uncertainty of the
    high SNR template: same SNR of the target in magenta, 2 times the
    SNR of the target in green, 3 times in red and 5 times better in
    blue. Both plots are for the isotropic case, $\sigma_{\rm Q} =
    \sigma_{\rm U}$.}
  \label{fig:residual_bias_nocorr}
\end{figure}

\begin{figure}
  \centering
  \includegraphics[angle=0,width=0.48\textwidth]{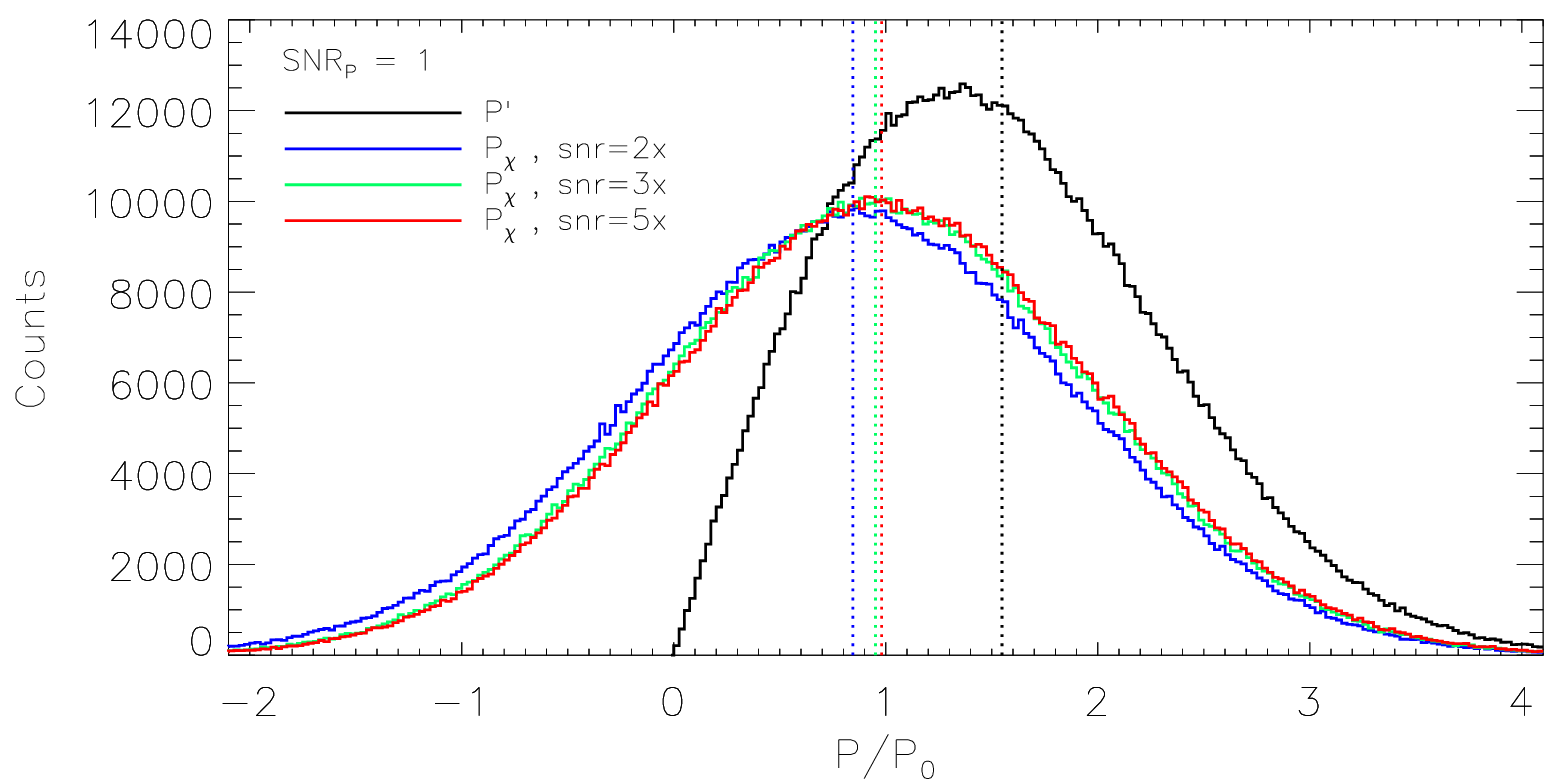}
  \includegraphics[angle=0,width=0.48\textwidth]{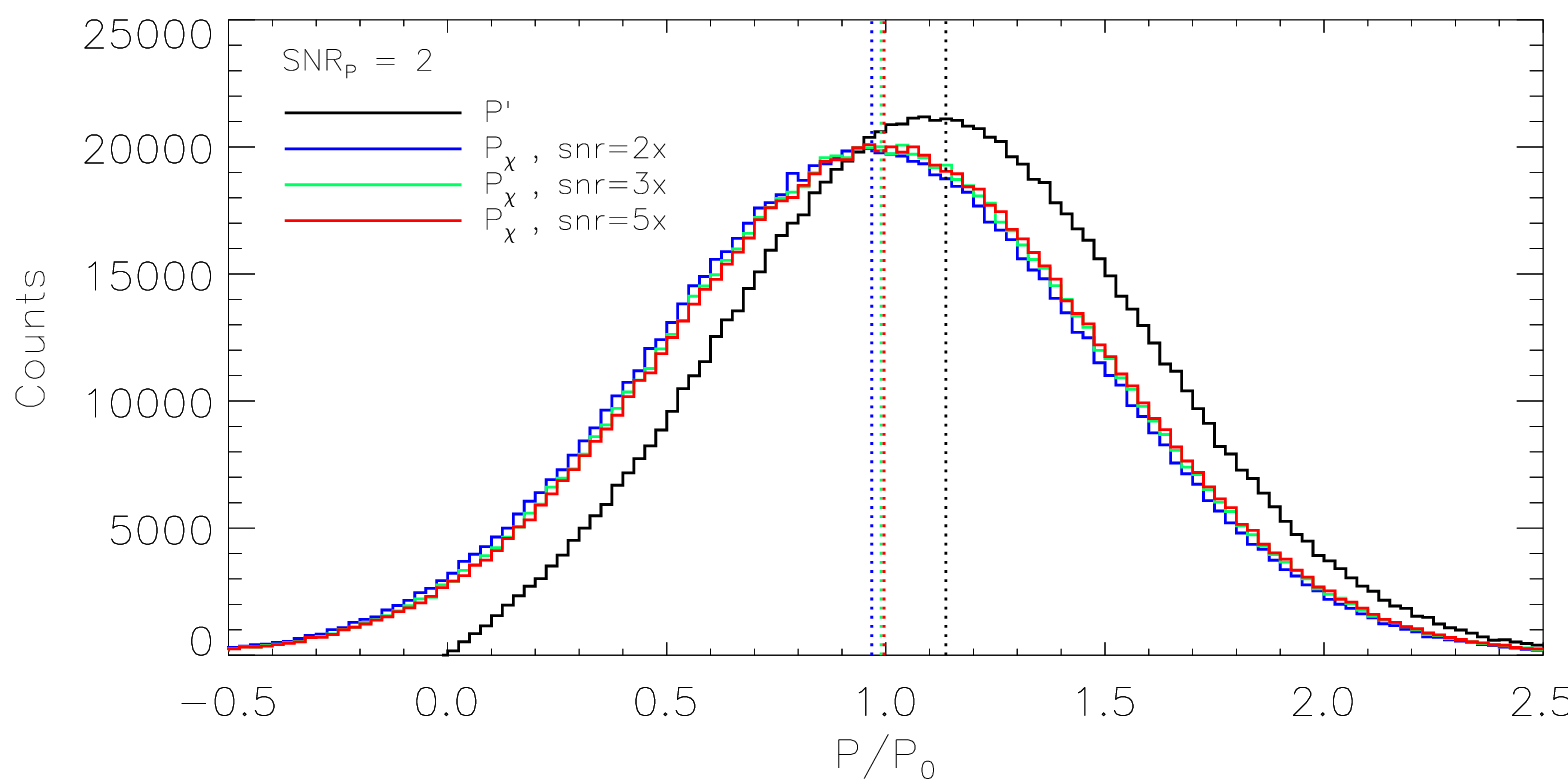}
  \includegraphics[angle=0,width=0.48\textwidth]{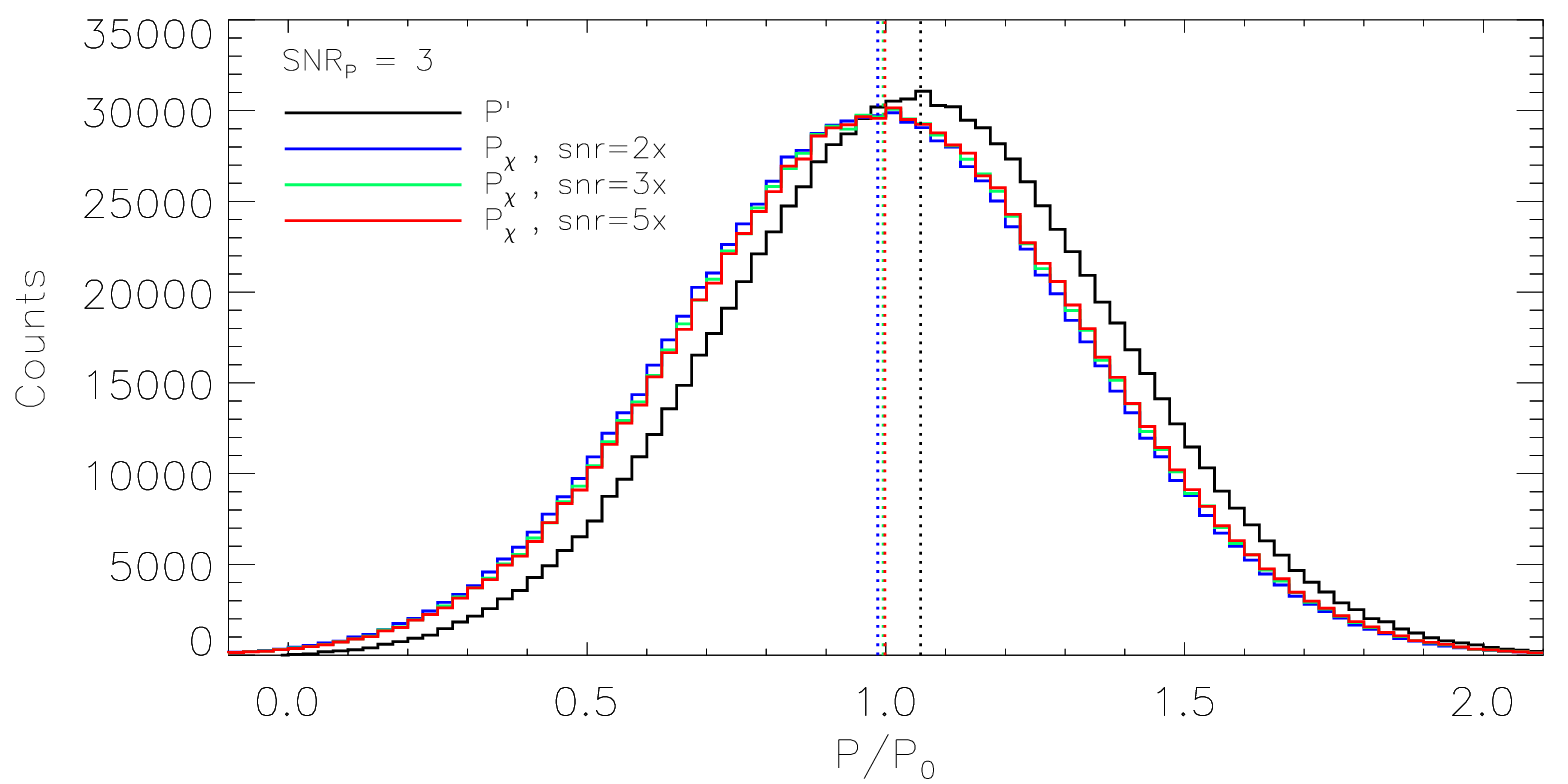}
  \caption{Histograms showing the distribution of $\pka$ (coloured
    lines) and the naive (black line) estimator for three values of
    SNR in the target (top to bottom) and three values of the SNR of
    the template (as a multiple of the SNR of the target). All
    histograms are for the isotropic case, $\sigma_{\rm Q} =
    \sigma_{\rm U}$. }
  \label{fig:dists}
\end{figure}

Using Monte Carlo simulations, we measure the residual bias in a
single pixel for four polarisation amplitude estimators:
\begin{enumerate}
\item $\pnaive=P' = \sqrt{Q'^2+U'^2}$, the naive estimator with no
  correction for bias.
\item $\pmas$, the Plaszczynski et al. modified asymptotic estimator, from
Eq.\,\ref{eq:p_mas}. 

\item $\pka$, the known-angle estimator, from Eq.\,\ref{eq:pdeb_chi}.
\end{enumerate}

\subsection{Isotropic case}

We ran $10^6$ Gaussian noise realisations for the $Q,U$ Stokes
parameters. For the known-angle estimator, we also produced an
additional set of $10^6$ noise realisations for the high
signal-to-noise template $Q,U$, which accounts for the uncertainty in
the known angle. We start with the simpler case where $\sigma_Q$ and
$\sigma_U$ are not correlated.  We studied the bias, defined as ${\rm
  bias} = \hat{p}-P_0$ and the risk function $=\langle(\hat{p}-P_0)^2
\rangle$, for the different estimators.  In
Fig. \ref{fig:residual_bias_nocorr} we show on the top panel the
residual bias of the estimators as a function of the observed
SNR$_{\rm P} \equiv P_0/\sigma_P$, with $\sigma_P$ defined in
Eq. \ref{eq:sigma_P}. We show four lines for $\pka$, which correspond
to different values of the SNR of the angle template.  The bottom
panel of Fig. \ref{fig:residual_bias_nocorr} shows the risk function
for the different estimators.

We can see that the known-angle estimator performs very well in the
case when the polarisation angle is known with relatively high
precision, i.e. when the SNR of the template is $>2$--3 times the SNR of
the target.

The construction of the known-angle estimator allows for negative
values of $\pka$. Fig. \ref{fig:dists} shows the distributions of
$\pka$ simulations for three different values of the SNR of the target
and the template.

\subsection{General case}
\label{sec:general}

\begin{figure*}
  \centering 
  \newcommand{\widthfig}{1}
  \includegraphics[angle=0,width=\widthfig\textwidth]{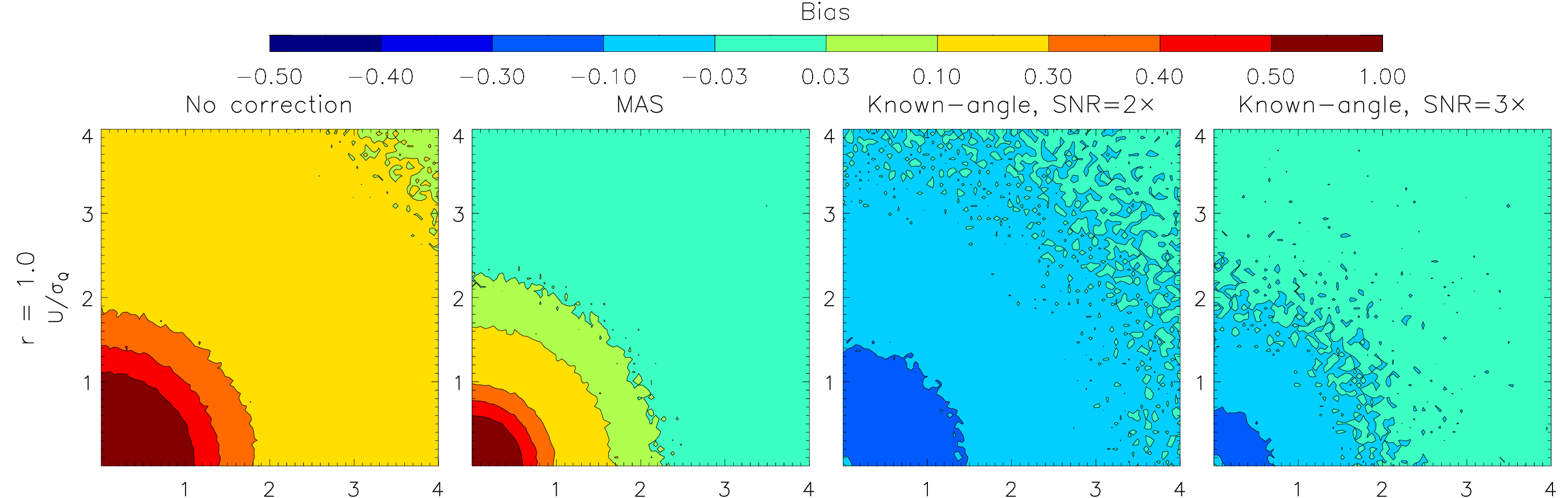}
  \includegraphics[angle=0,width=\widthfig\textwidth]{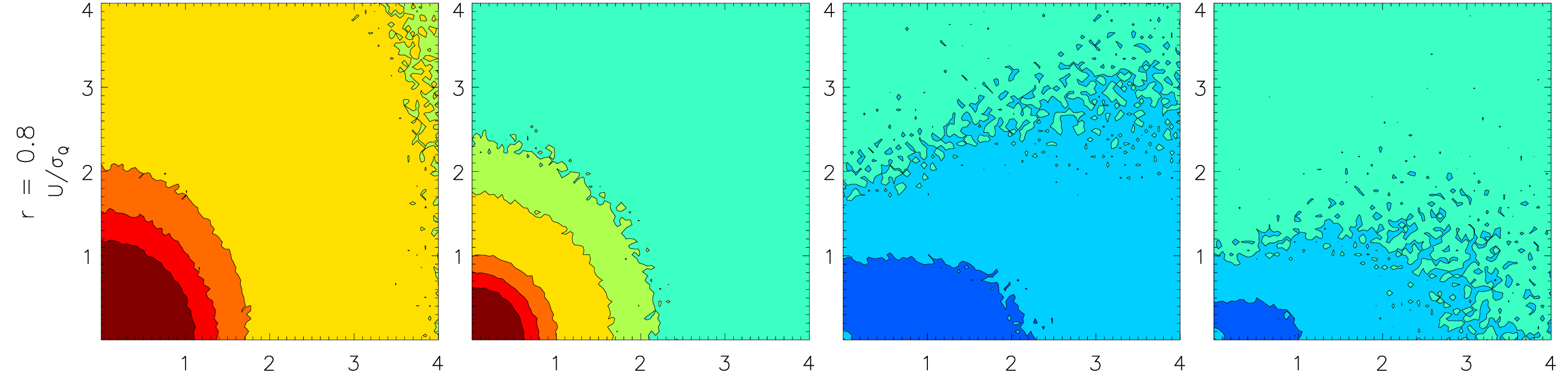}
  \includegraphics[angle=0,width=\widthfig\textwidth]{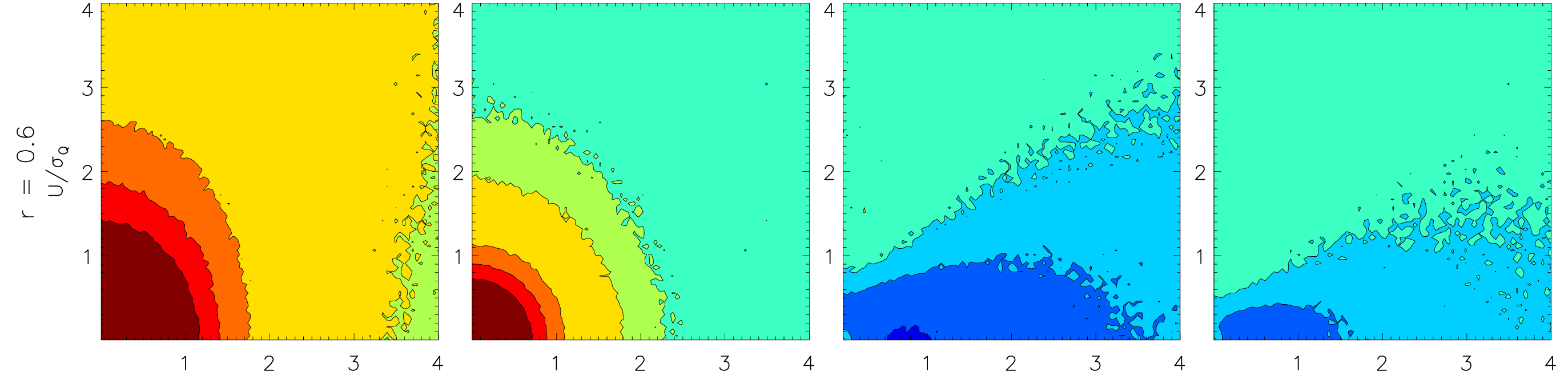}
  \includegraphics[angle=0,width=\widthfig\textwidth]{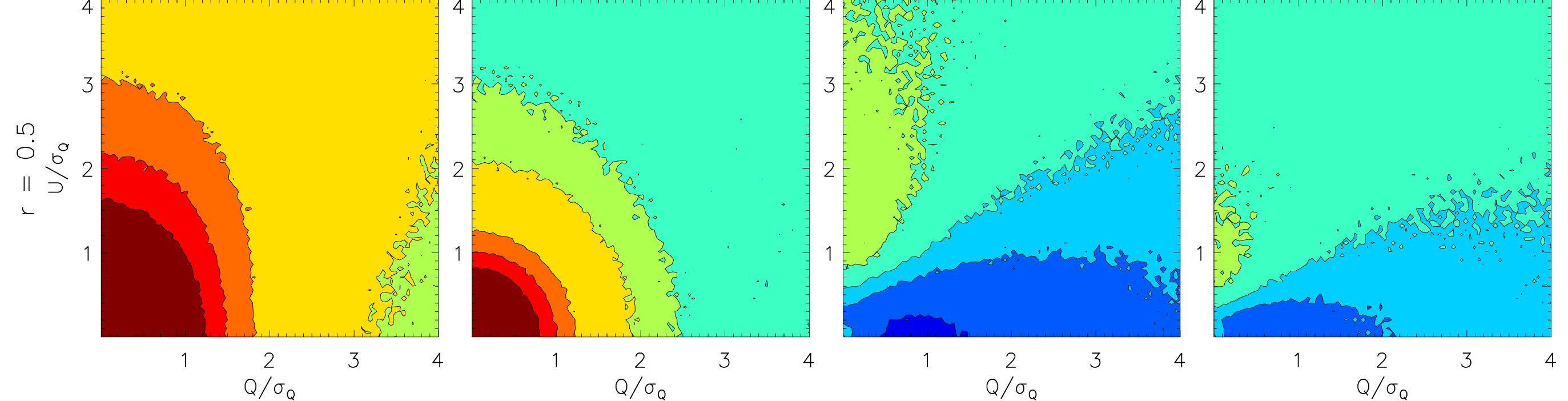}
  \caption{Contours of the mean residual bias of three estimators for
    the polarised intensity as a function of $(Q,U)$, calculated using
    Monte Carlo simulations for four different values of the error
    ellipse axial ratio, $r$.  We take $\rm{cov_{QU}}=0$ (there is no
    loss of generality, as the covariance can always be eliminated by
    a rotation of the $Q-U$ axes). The colour scale represents the
    percentage bias of the estimated polarisation value in each
    pixel. The {\it left} column shows the naive estimator,
    $\pnaive$. The second column shows the MAS estimator, $\pmas$,
    from \citet{plaszczynski:14}. The third and forth columns show the
    known-angle estimator, $\pka$, presented in Eq. \ref{eq:pdeb_chi}
    for two different SNR levels of the angle template: 2 times and 3
    times the SNR of the target. }
  \label{fig:bias_contours}
\end{figure*}

Here we allow for asymmetric errors in $Q,U$, i.e. for $r<1$. We
created a grid of $100\times 100$ different values of $Q_0/\sigma_Q$
and $U_0/\sigma_Q$, in the range $Q_0/\sigma_Q \le 4$, $ U_0/\sigma_Q
\le 4$, using a uniform spacing. This was repeated for four values of
$r$.  We set $\theta=0$, so that $r = \sigma_U/\sigma_Q$ (the pattern
in the $(Q,U)$ plane simply rotates for other values of $\theta$).
Then, $\times10^5$ Gaussian noise realisations are added to the each
point in the grid.  We calculated the `observed' polarisation
amplitude $P'$ from the noisy simulations, applied the four estimators
to each simulation, and in each case measured the mean bias $b =
(\left<\hat{p}\right> - P_0)$ for each pixel.  The first column of
Fig.~\ref{fig:bias_contours} shows the bias of the naive estimator
$P'$.  The second column shows the residual bias after using the
modified asymptotic estimator (MAS). The biased regions in the SNR
plane reduces considerably in comparison with the first column that
has no correction.  The third and forth columns shows the residual
bias using the known-angle estimator. As this estimator requires an
independent value for the polarisation angle along with the observed
values $(Q',U')$, we generated an additional $\times 10^5$ Gaussian
realisations for the $Q,U$ Stokes parameters of the template and with
them, we reconstructed the known polarisation angle, centred at the
true value, $\chi_0$, for each SNR value. Here, the SNR of the
template is 2 (third column) and 3 (forth column) times the SNR of the
target.

As expected, apart from numerical noise these results agree with the
exact calculation shown in Fig.~\ref{fig:residual_bias}.

\subsection{\wmap simulations}
\label{sec:wmap_sims}
\wmap provided maps at five frequency bands between 23 and 94\,GHz
\citep{bennett:13}. In the lower three frequencies the sky
polarisation is dominated by Galactic synchrotron emission, with a
brightness temperature that drops steeply with frequency ($T_\nu
\propto \nu^{\beta}$, with $\beta \approx -3$).  Since the brightness
temperature sensitivity is similar in all bands, the highest SNR is at
23\,GHz (K band), where large areas of diffuse polarised emission have
$\mbox{SNR}>3$ after smoothing to 1\dg\ FWHM resolution. In these
bands the synchrotron polarisation angle reflects the Galactic
magnetic field direction in the source regions, and is expected to be
almost independent of frequency. The most likely cause of any
frequency variation is superposition on the line of sight of regions
with different field directions and also difference spectral indices
$\beta$; however, the variation of $\beta$ for the synchrotron
component is small as has been shown by \citet{Fuskeland2015} and
\citep{Vidal2015}. On the other hand, the higher \wmap bands
begin to be sensitive to dust polarisation, which has $\beta \approx
+1.7$, and \planck data confirm that this is generally significantly
misaligned with synchrotron \citep{PIP_XIX,PIP_XXII}.  For this reason
we only consider the three lowest \wmap bands below.

\begin{figure}
  \centering
  \hspace{-0.5cm} \includegraphics[angle=0,width=0.5\textwidth]{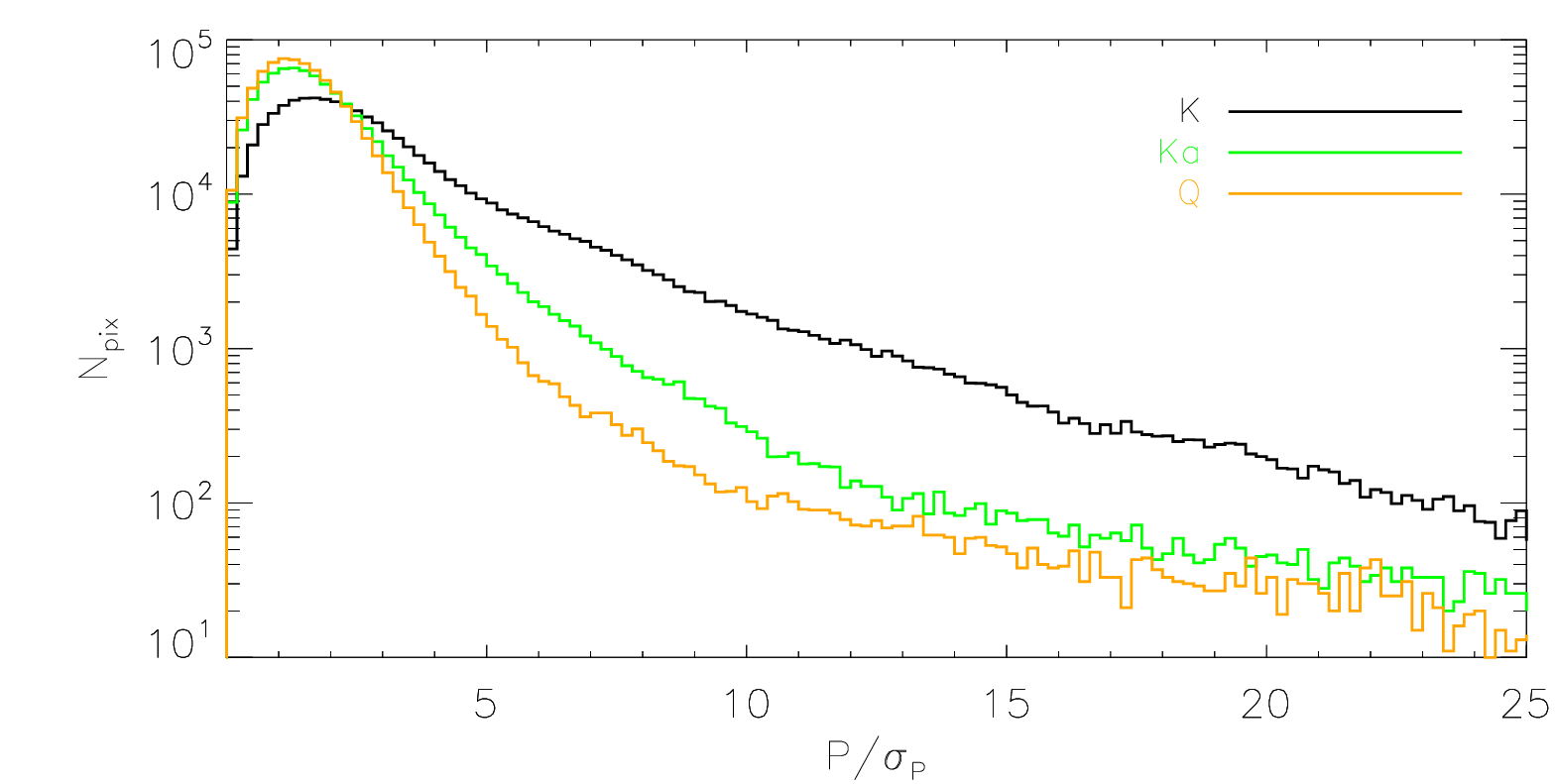}
  \caption{Histograms of the estimated signal-to-noise ratio of the
    polarisation amplitude, SNR$_{\rm P} =P/\sigma_P$ of \wmap data
    in 3 frequency bands. The histograms are made using the full sky
    maps at an angular resolution of 1\dg, with $\nside = 256$. The
    percentage of pixels where the SNR at K-band is larger than the
    SNR at Ka and Q bands is 75\% and 78\% respectively. }
  \label{fig:wmap_snr_pol}
\end{figure}

\begin{figure}
  \centering
  \hspace{-0.5cm} \includegraphics[angle=0,width=0.5\textwidth]{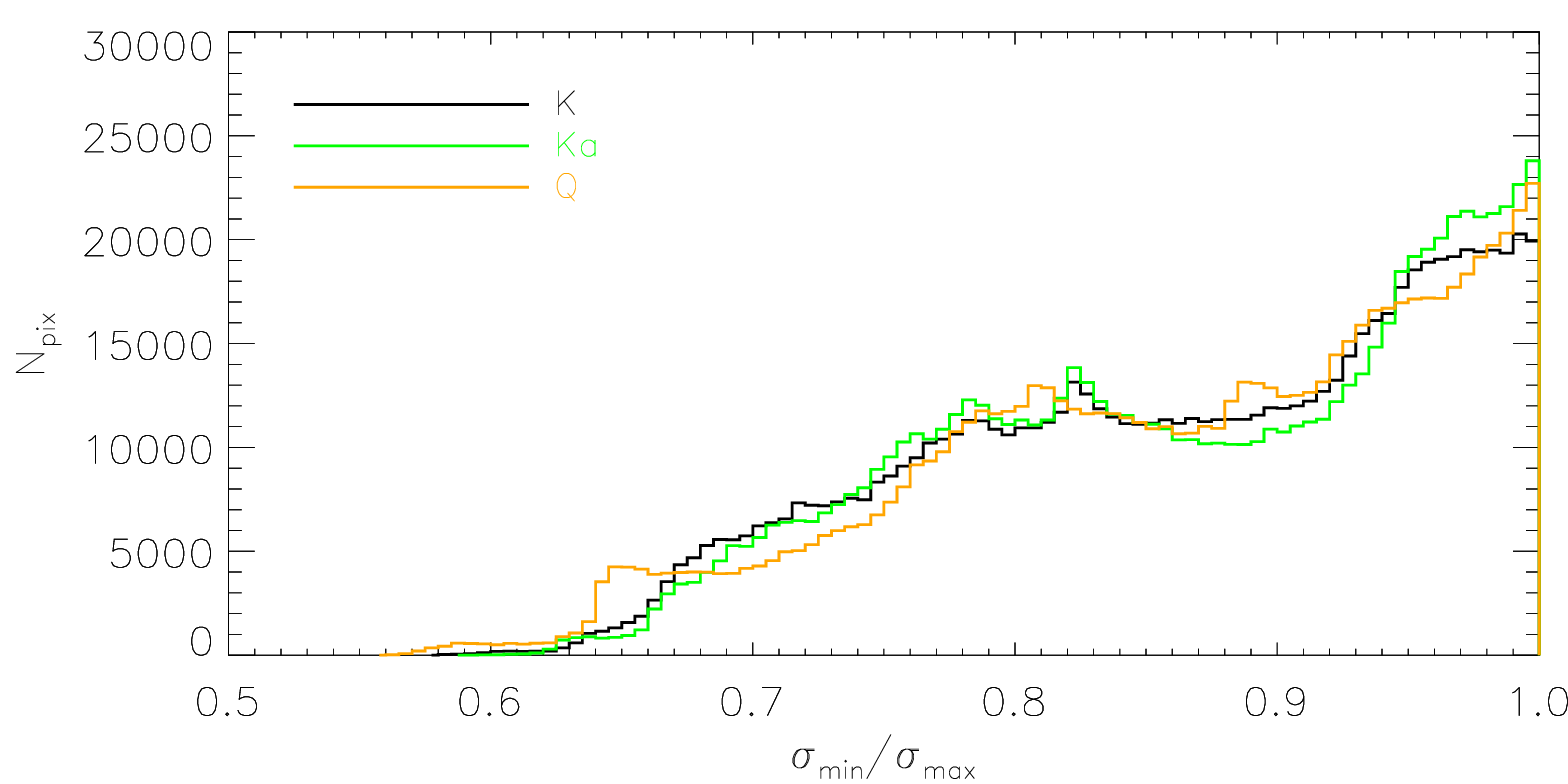}
 \caption{Histograms of the polarisation error ellipse axial ratio
   (Eq. \ref{eq:r}) of \wmap data in 3 frequency bands. The histograms
   are made using the full sky maps at an angular resolution of 1\dg,
   with $\nside=256$.}
  \label{fig:ratio_wmap}
\end{figure}

Fig.~\ref{fig:wmap_snr_pol} show histograms of the SNR for the Stokes
parameters $(Q,U)$ of the three frequency bands: K- (23\,GHz), Ka-
(33\,GHz) and Q- (41\,GHz) bands.  The polarisation SNR in K-band is
larger than the SNR in the other bands for almost the entire sky; in
fact, the SNR in the Ka-, and Q-bands rarely exceeds 3. In
Fig.~\ref{fig:ratio_wmap} we show histograms of the axial ratio of the
error ellipses for all the pixels in these \wmap bands. The mean ratio
for all the bands is $r= 0.86$.

\begin{figure}
  \centering 
  \newcommand{\widthfig}{0.49}
  \newcommand{\anglefig}{90}
  \includegraphics[angle=\anglefig,width=\widthfig\textwidth]{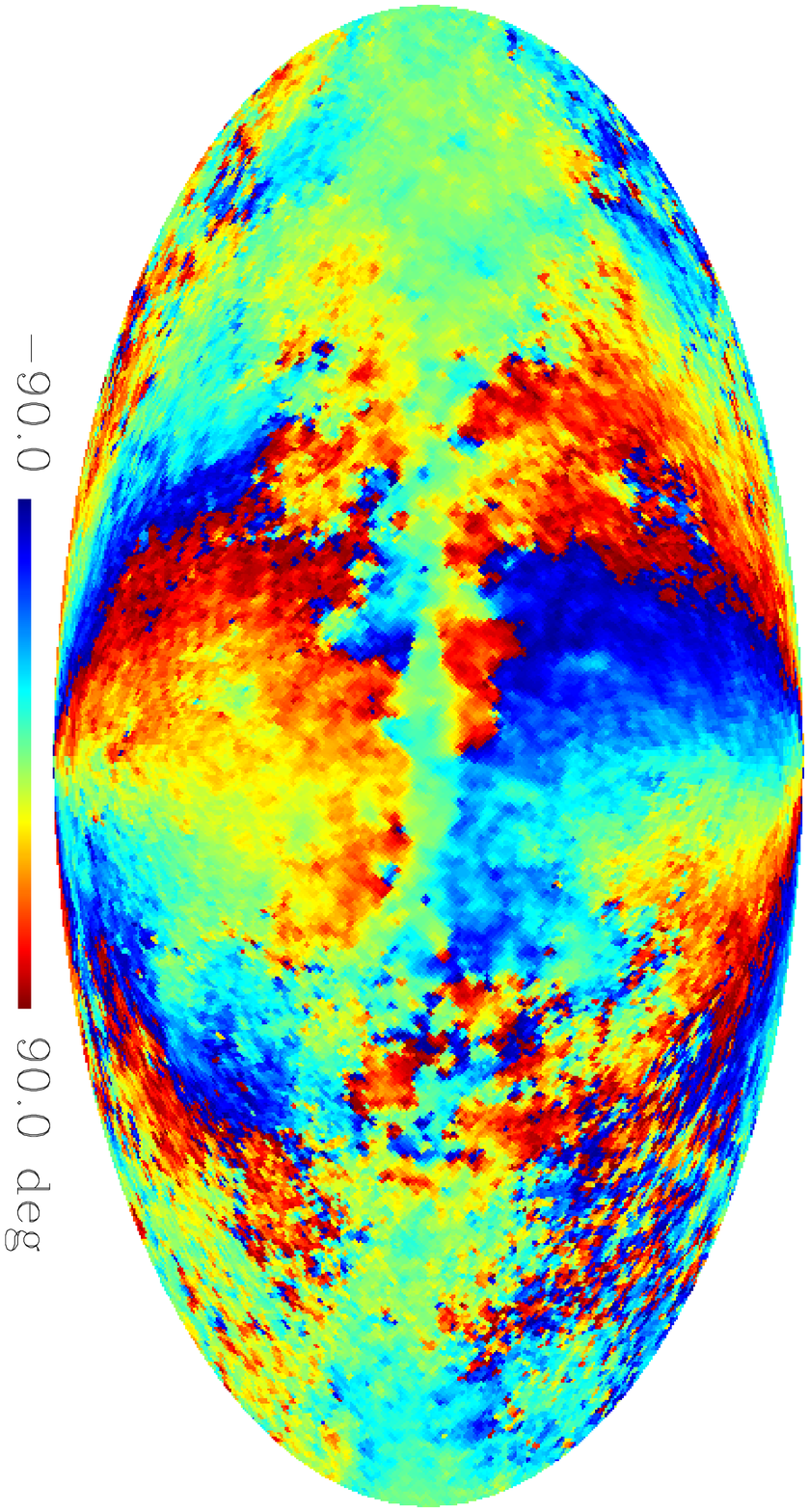}
  \includegraphics[angle=\anglefig,width=\widthfig\textwidth]{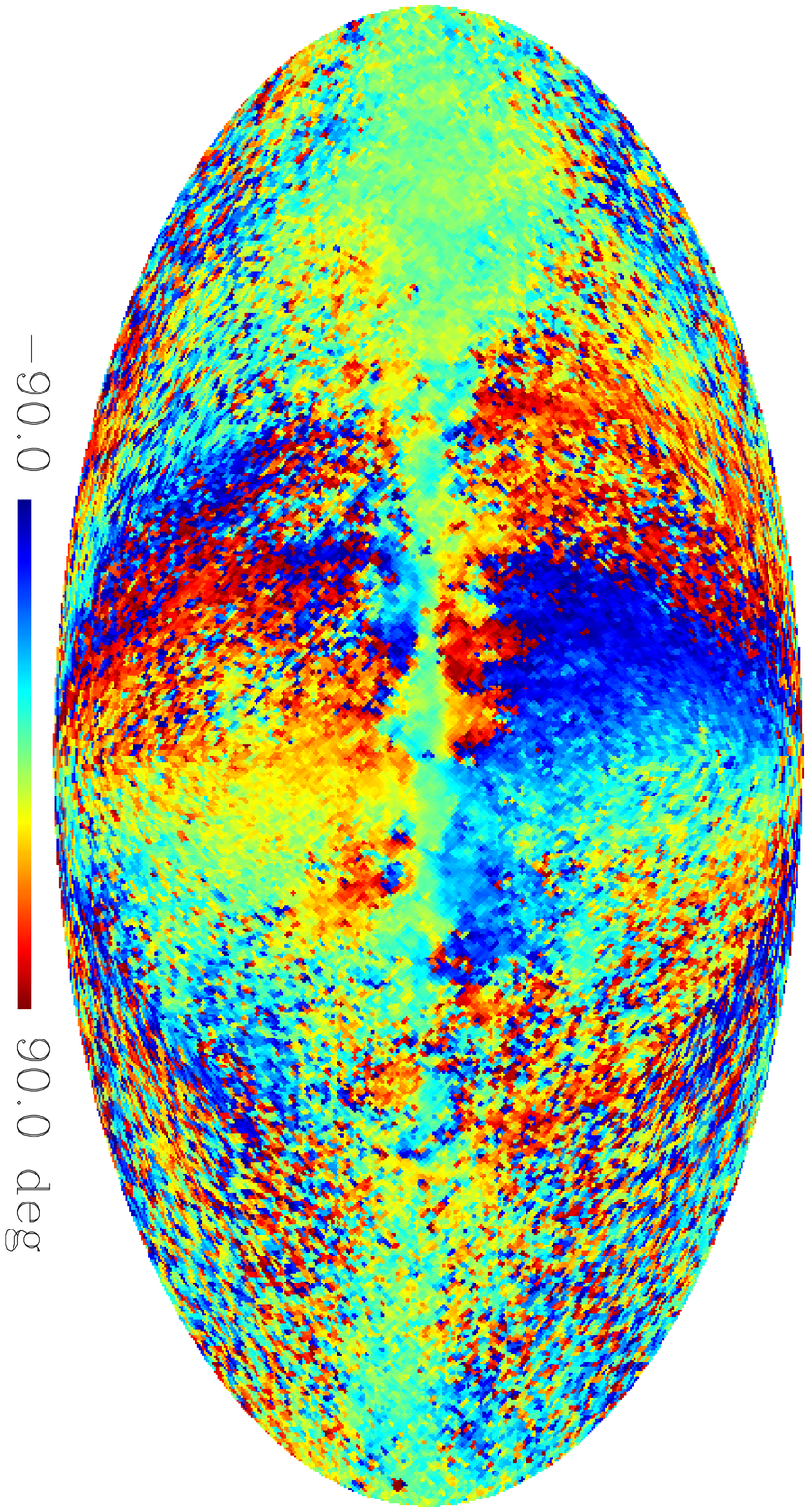}
  \includegraphics[angle=\anglefig,width=\widthfig\textwidth]{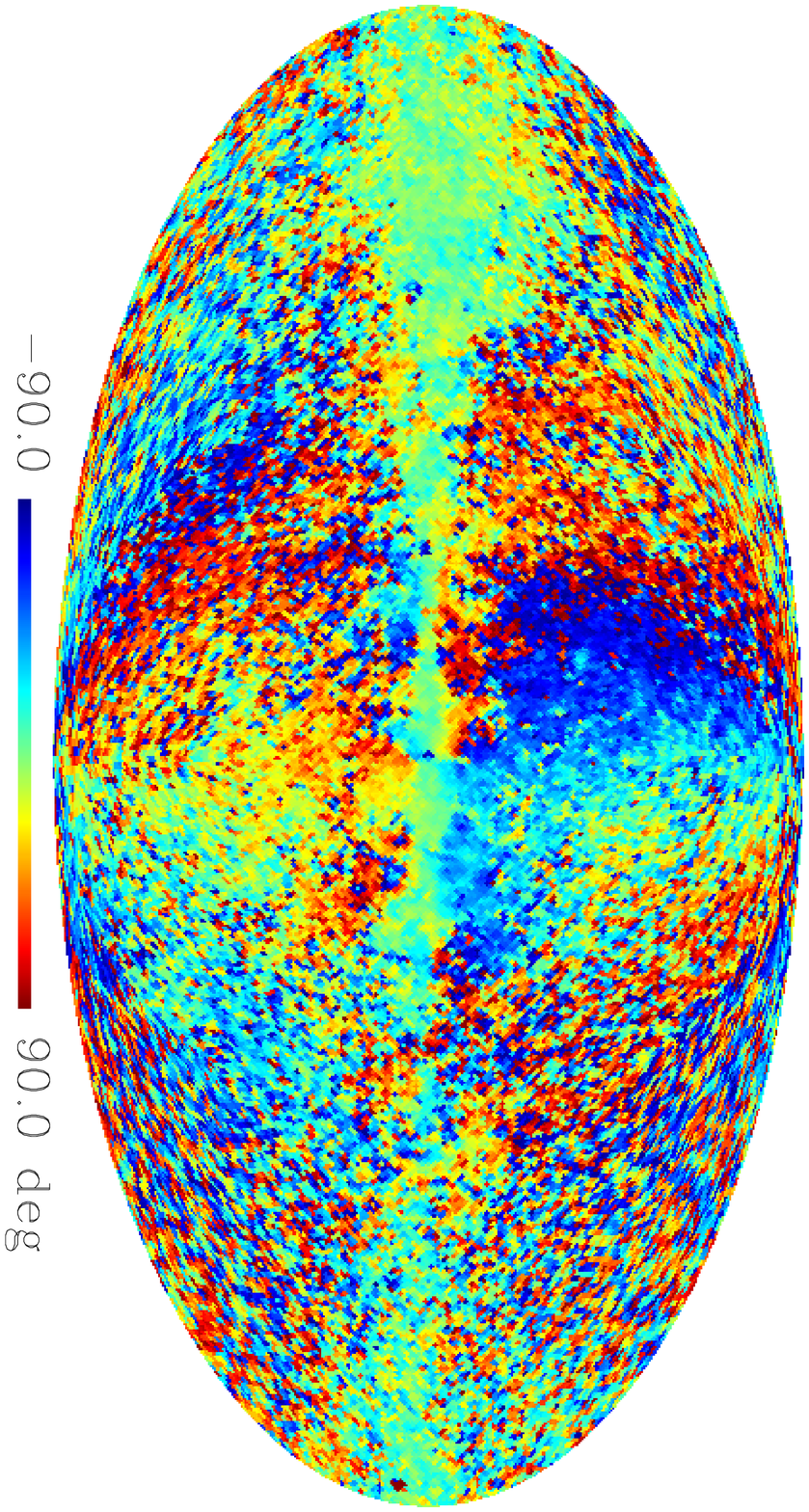}
  \caption{Full sky polarisation angle map. On {\it top} is the
    reconstructed angle map from the Planck Sky model with the added
    noise from \wmap K-band data. The {\it middle} and {\it bottom}
    panels show the observed polarisation angle from \wmap Ka and
    Q-bands respectively.  }
  \label{fig:angles_wmap}
\end{figure}

We used the \planck Sky Model (PSM) \citep{Delabrouille2014} to
simulate the polarised sky at K-, Ka- and Q-bands at an angular
resolution of 1\dg, from which we can obtain maps for the unbiased
polarisation amplitude, $P_0$. The model is closely based on the \wmap
K-band map, and uses realistic spectral indices.  We added random
noise, generated using the \wmap covariance matrices $C$ for each
pixel, to the simulated Stokes $Q$ and $U$ map. The maps are generated
at HEALPix $\nside=512$. Smoothing was done by transforming to
spherical harmonics, which were then divided by the \wmap instrumental
window functions and multiplied by the window function of a Gaussian
beam with 1\dg\ FWHM, then re-transformed. Noise and sky model maps
were downgraded to $\nside = 64$ to give approximately independent
pixels, and summed to give maps of the `observed' polarisation
amplitude, $P'$. In the top panel of Fig. \ref{fig:angles_wmap} we
show the reconstructed simulated polarisation angle map at K-band. The
middle and lower panel of Fig. \ref{fig:angles_wmap} show the observed
polarisation angle by \wmap at Ka and Q-bands respectively. We can see
that the simulated K-band angles, which we use as our template is very
similar (but less noisy than) the observed angle at Ka and Q
band. This is the characteristic that the known-angle estimator
requires.

We corrected for the bias in these $P'$ maps using the same three
estimators discussed in Section~\ref{sec:general}.  Ka- and Q-bands
are corrected using the known-angle estimator where the angle
information required by the $\pka$ estimator is measured from the
K-band map. We then compared these de-biased maps with the true
polarisation amplitude map from the PSM simulations.  In Table
\ref{tab:delta_p_deb} we list the normalised bias value, averaged over
the entire sky of 500 simulations for the three frequency bands that
we studied.

\renewcommand{\arraystretch}{1.1}
\begin{table}
  \caption{Full-sky averaged values for the normalized polarisation
    bias and standard deviation of each estimator in simulated \wmap data,
    over 500 simulations. See Fig. \ref{fig:bias_histograms} for the 
    histograms of the errors $\hat{p} - P_0$.} 
\centering  \vspace{0.5cm}
  \begin{tabular}{lccrrrr}
    \toprule
    Estimator & \multicolumn{2}{c}{\kband}       & \multicolumn{2}{c}{\kaband}    & \multicolumn{2}{c}{\qband} \\
       &$\left<\Delta  \hat{p} \right>^a$ & std$[\Delta\hat{p}]^b$ &  
    $\left<\Delta  \hat{p} \right>^a$ & std$[\Delta\hat{p}]^b$ &
    $\left<\Delta  \hat{p} \right>^a$ & std$[\Delta\hat{p}]^b$ \\
    \midrule
    $\Delta \pnaive$   & 0.20 & 0.98 &  0.50  &  0.91  &  0.66  &  0.87 \\
    $\Delta \pmas  $     & 0.06 & 1.00 &  0.27  &  0.93  &  0.42  &  0.88 \\
    $\Delta \pka   $   &  --  &  --  & $-0.05$ &  1.01  & $ -0.03$ &  1.01 \\
    \bottomrule
  \end{tabular}
\begin{flushleft}
$^a$ mean normalized bias, $\langle \hat{p} - P_0\rangle/\langle
  \sigma_P\rangle$ where $\sigma_P$ is given by Eq.~\ref{eq:sigma_P}.

$^b$ Standard deviation of $\hat{p} - P_0$, again in units of the mean error. 

\end{flushleft}
   \label{tab:delta_p_deb}   
\end{table}

\begin{figure}
  \centering 
  \newcommand{\widthfig}{0.48}
  \newcommand{\anglefig}{0}

  \includegraphics[angle=\anglefig,width=\widthfig\textwidth]{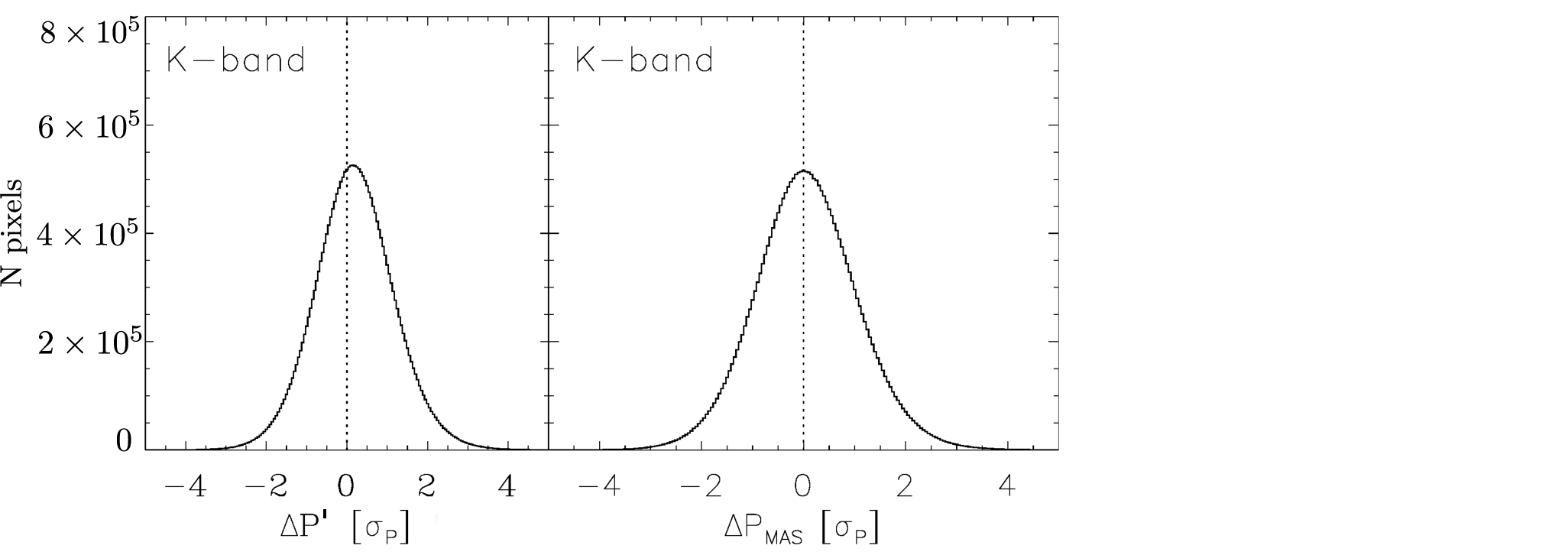}
  \includegraphics[angle=\anglefig,width=\widthfig\textwidth]{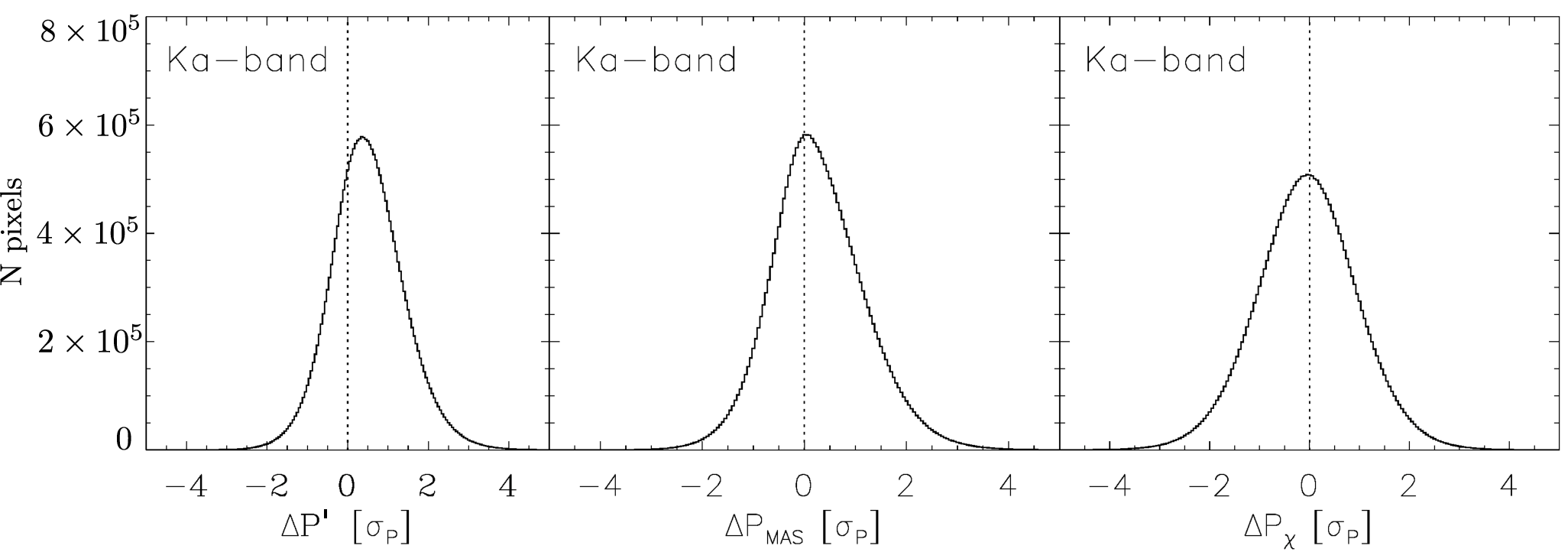}
  \includegraphics[angle=\anglefig,width=\widthfig\textwidth]{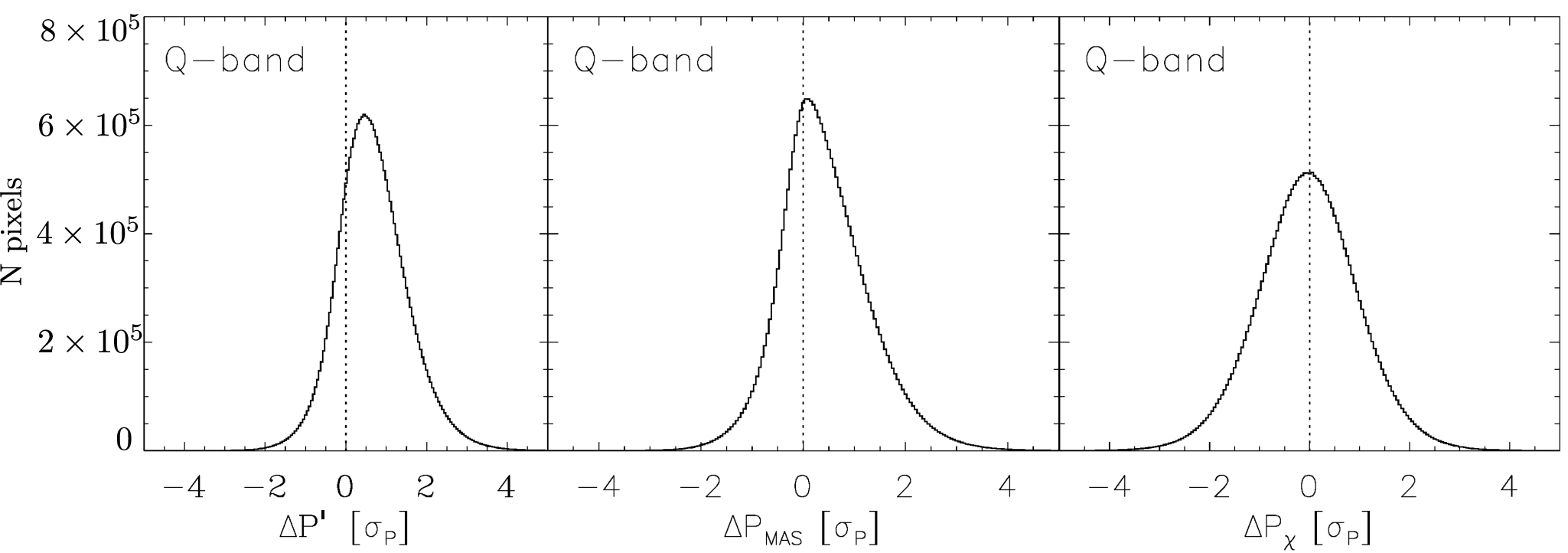}
  \caption{Histograms showing the bias of the three estimators studied
    for the simulated \wmap data at K, Ka and Q band described in the
    text. Histograms combine all 500 simulations and all sky
    pixels. They show the distribution of the difference between the
    estimators and the true polarisation amplitude $P_0$, in units of
    the average polarisation noise, $ \left< \sigma_{P} \right>$.  The
    column on the left shows the histograms of the polarisation bias
    of the naive estimator, i.e. the $\Delta P_0 = P' - P_0$ (no bias
    correction).  The second column shows the residual bias for the
    MAS estimator, $\Delta P_{\rm MAS} = \pmas - P_0$. For the noisier
    bands (Ka and Q), the histograms are not centred at zero, implying
    that there is additional residual bias on the corrected map. The
    third column shows the histograms produced using the $\pka$
    estimator, $\Delta P_{\chi} = \pka - P_0$. Here the bias
    correction works much better and the distributions for the three
    bands are centred at zero. Table\,\ref{tab:delta_p_deb} lists the
    central values and spreads for all these histograms.}
  \label{fig:bias_histograms}
\end{figure}

The mean uncorrected bias, $\left<\Delta \pnaive\right> =
\left<P'-P_{0} \right>$ increases with frequency due to the decrease
in SNR.  $\pmas$ substantially reduce the bias at K and Ka bands, but
Q band is so noisy that its impact is relatively modest.  A clearer
view of the effect of the estimators is revealed in the histograms of
normalised errors, $P' - P_0 /\langle \sigma_P\rangle$, shown in
Fig. \ref{fig:bias_histograms}.  In Ka- and Q-band the histograms of
the uncorrected polarisation maps are not centred at zero (the three
plots in the left column).  Where available, i.e. for Ka- and Q-bands,
the $\pka$ estimator performs dramatically better than $\pmas$. Note
also that the histograms of the $\pka$ estimator have a clear Gaussian
shape. This shows that the uncertainty on the estimator
(Eq. \ref{eq:error_pchi}) can be used safely.

\begin{figure*}
  \centering 
  \newcommand{\widthfig}{0.33}
  \newcommand{\anglefig}{90}
 
   \includegraphics[angle=90,width=\widthfig\textwidth]{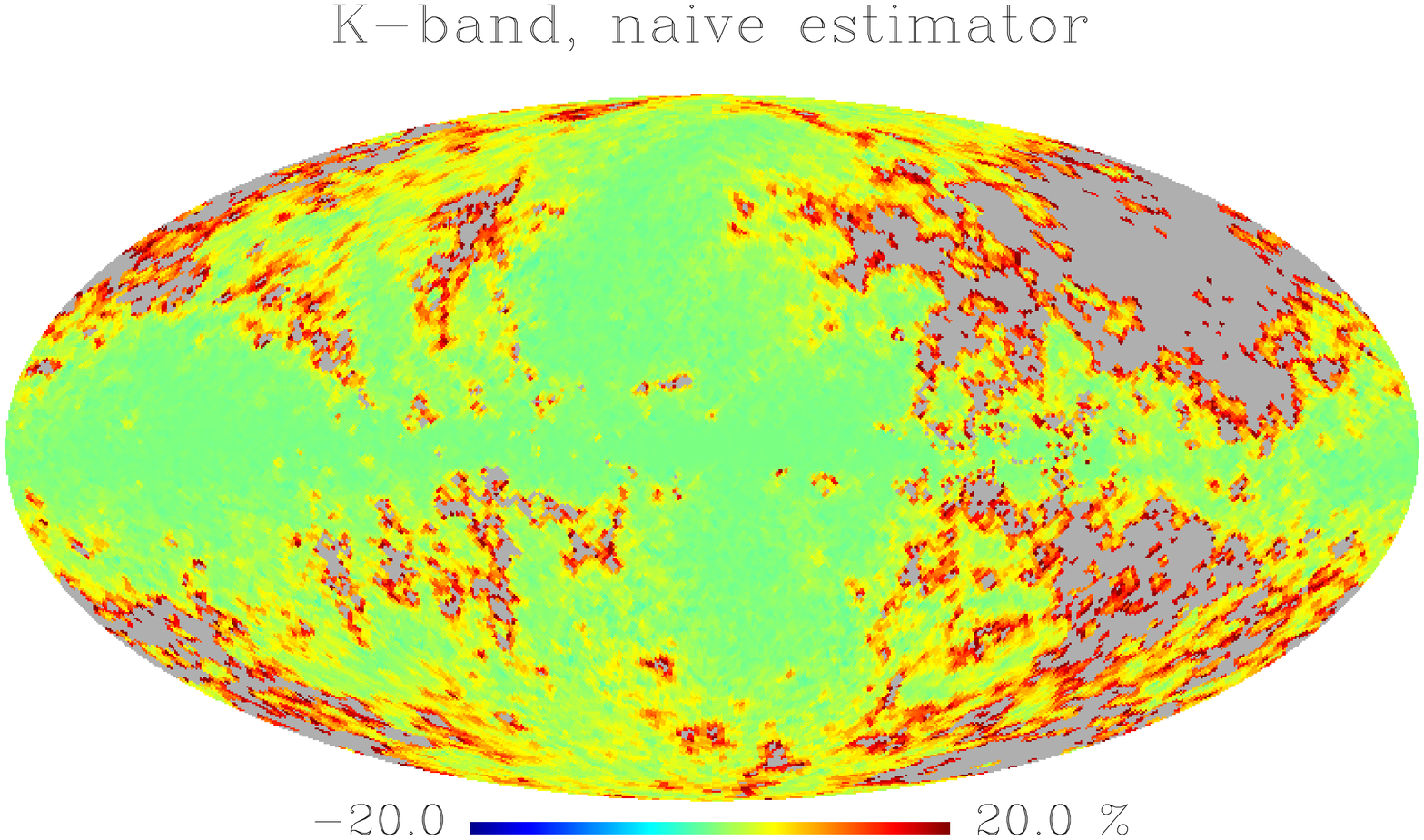}
   \includegraphics[angle=90,width=\widthfig\textwidth]{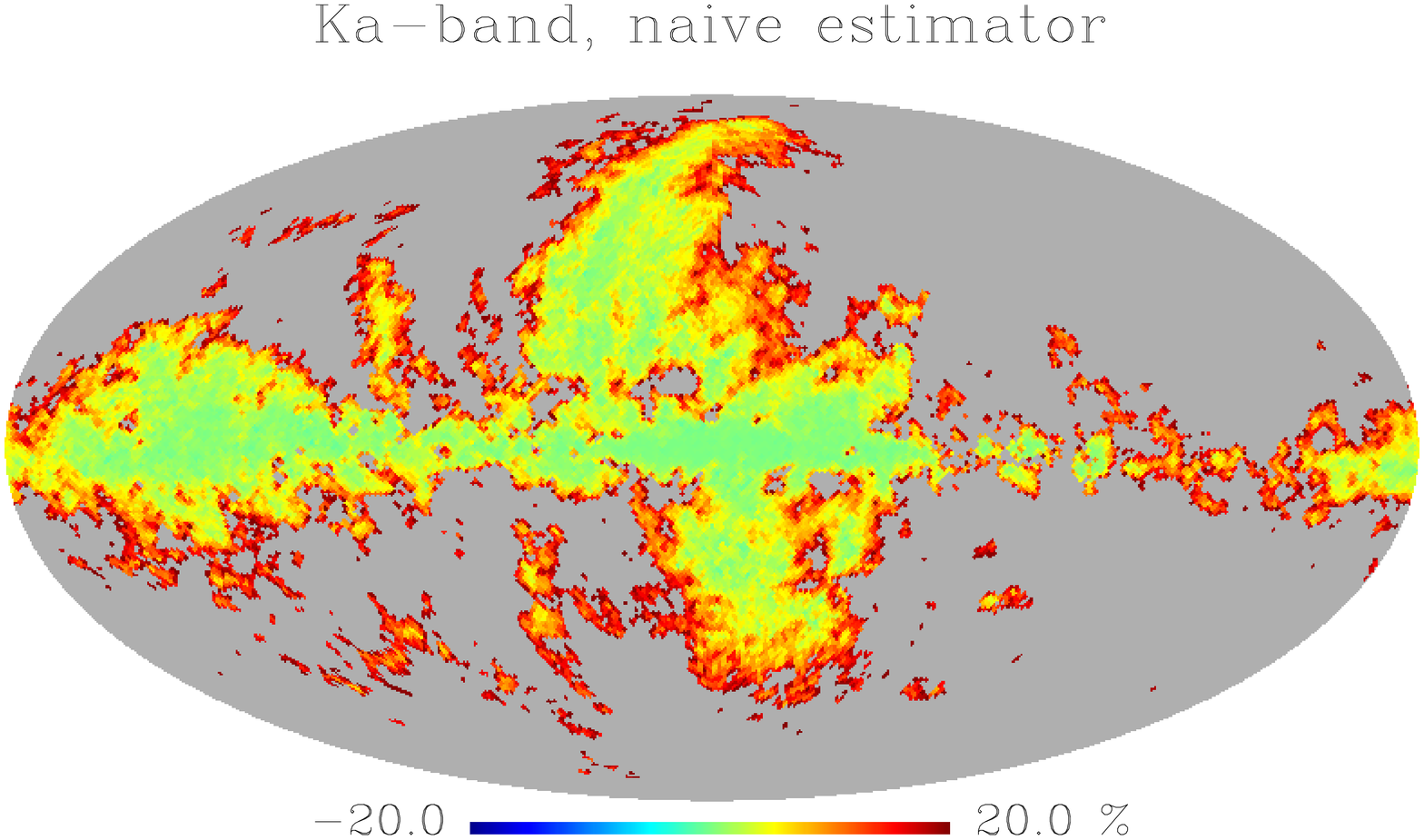}
   \includegraphics[angle=90,width=\widthfig\textwidth]{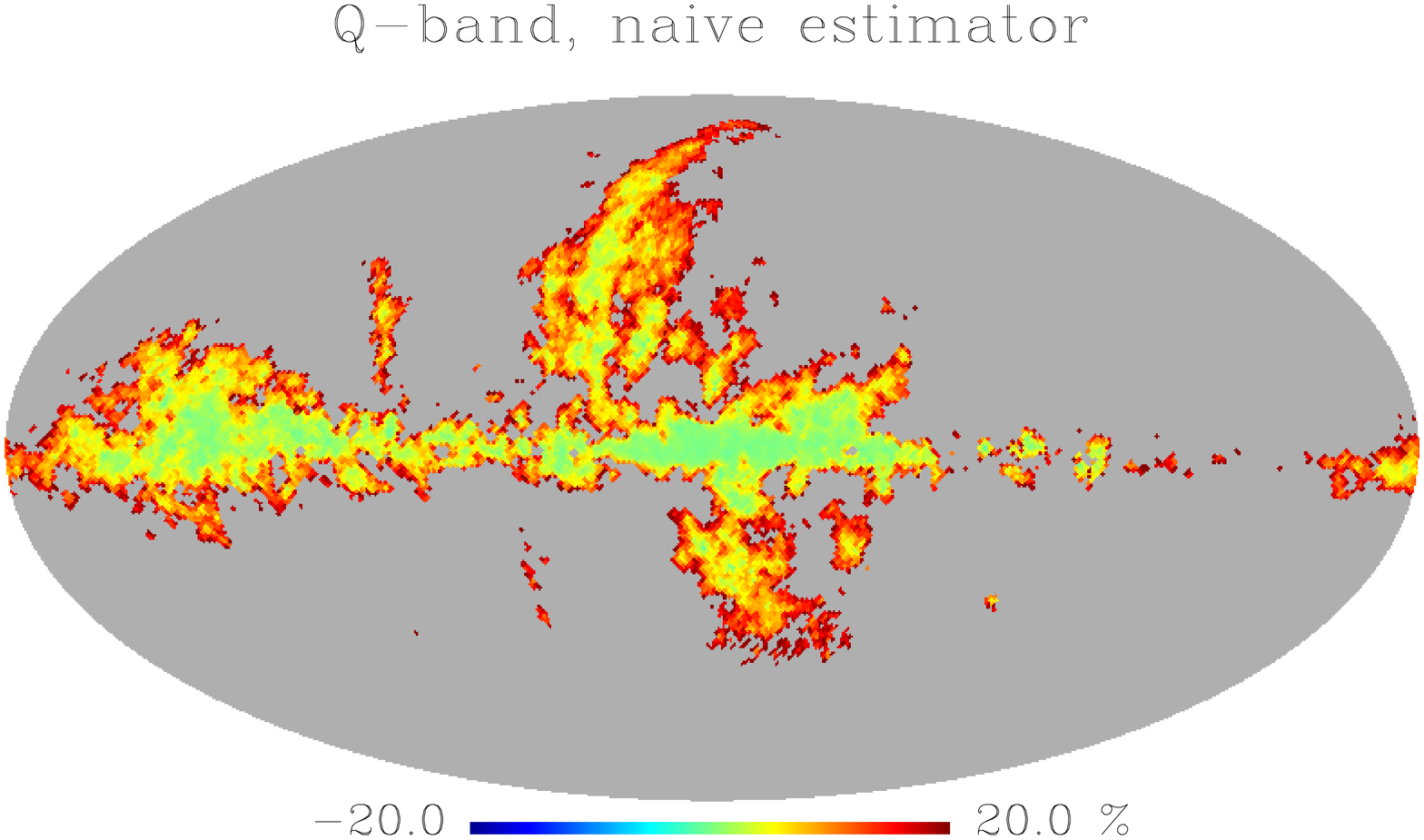}
   \includegraphics[angle=90,width=\widthfig\textwidth]{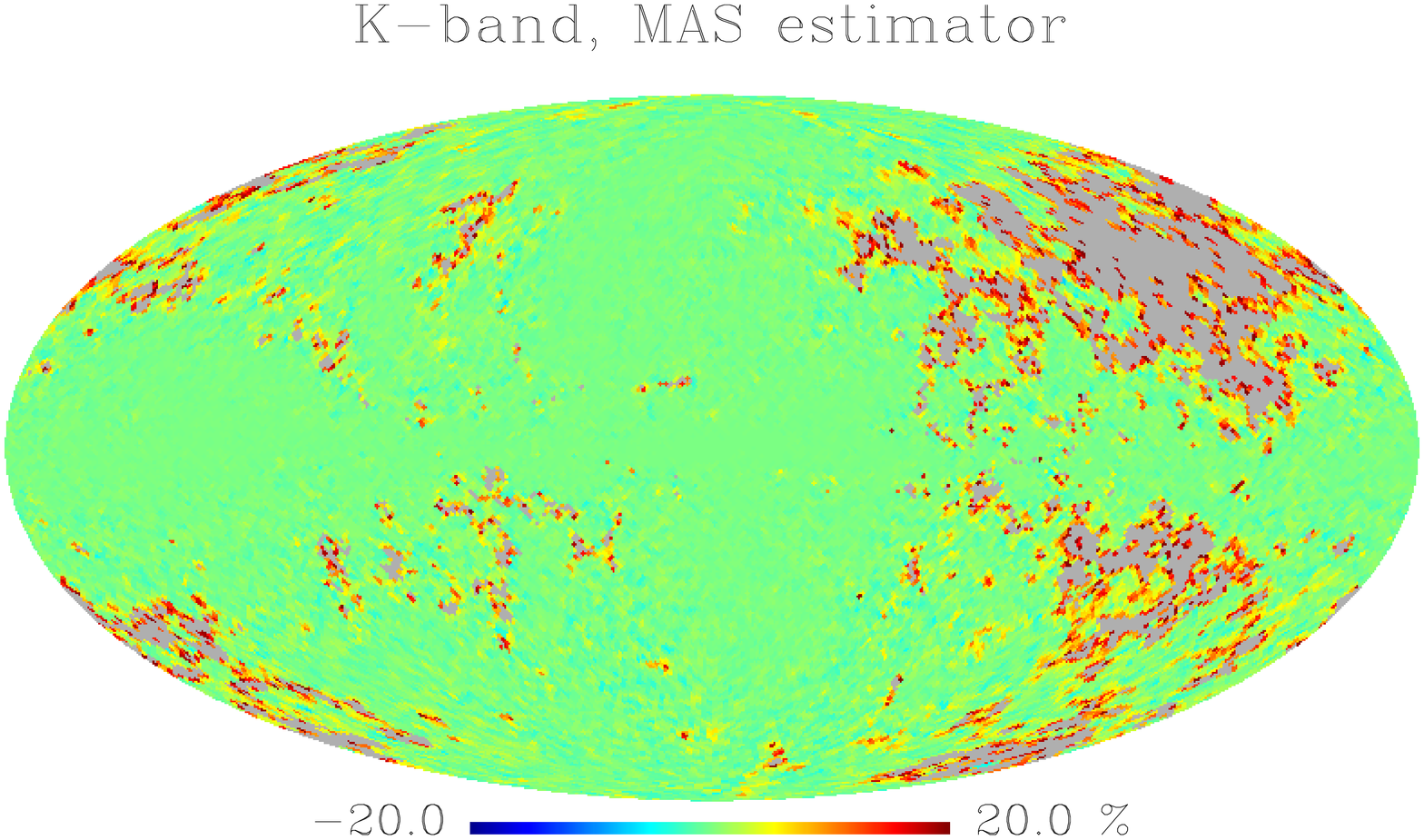}
   \includegraphics[angle=90,width=\widthfig\textwidth]{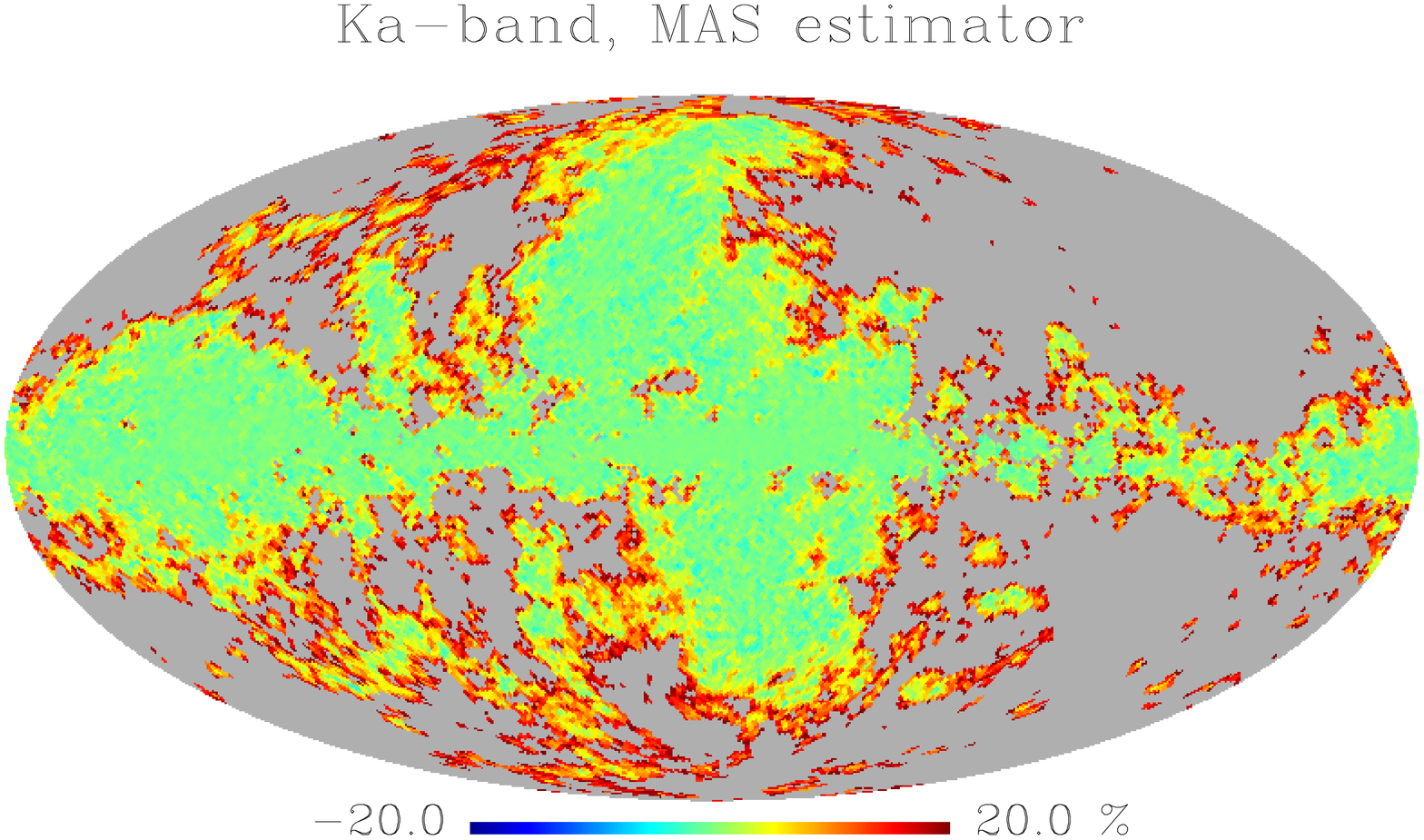}
   \includegraphics[angle=90,width=\widthfig\textwidth]{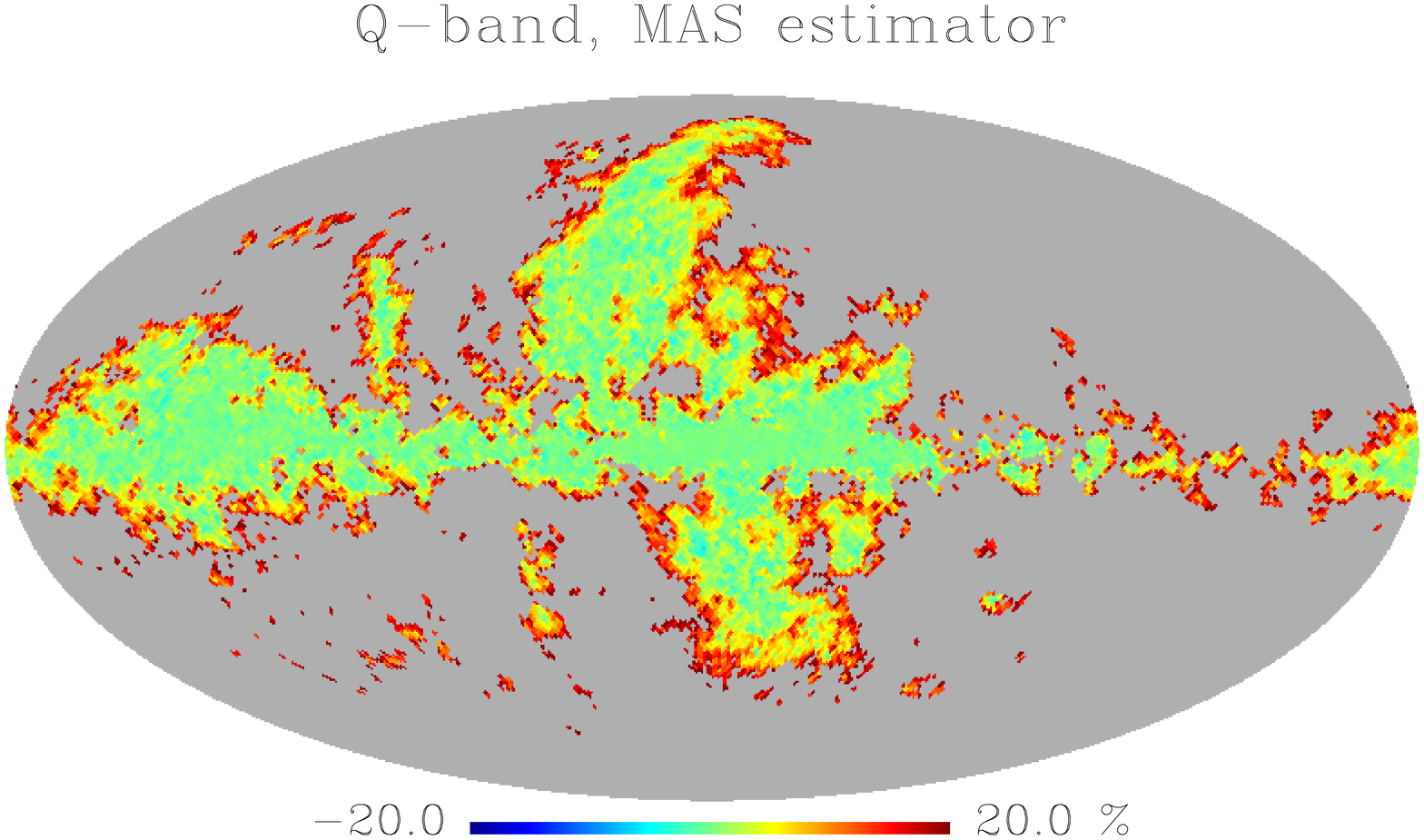}
   \raggedleft                                                   
   \includegraphics[angle=90,width=\widthfig\textwidth]{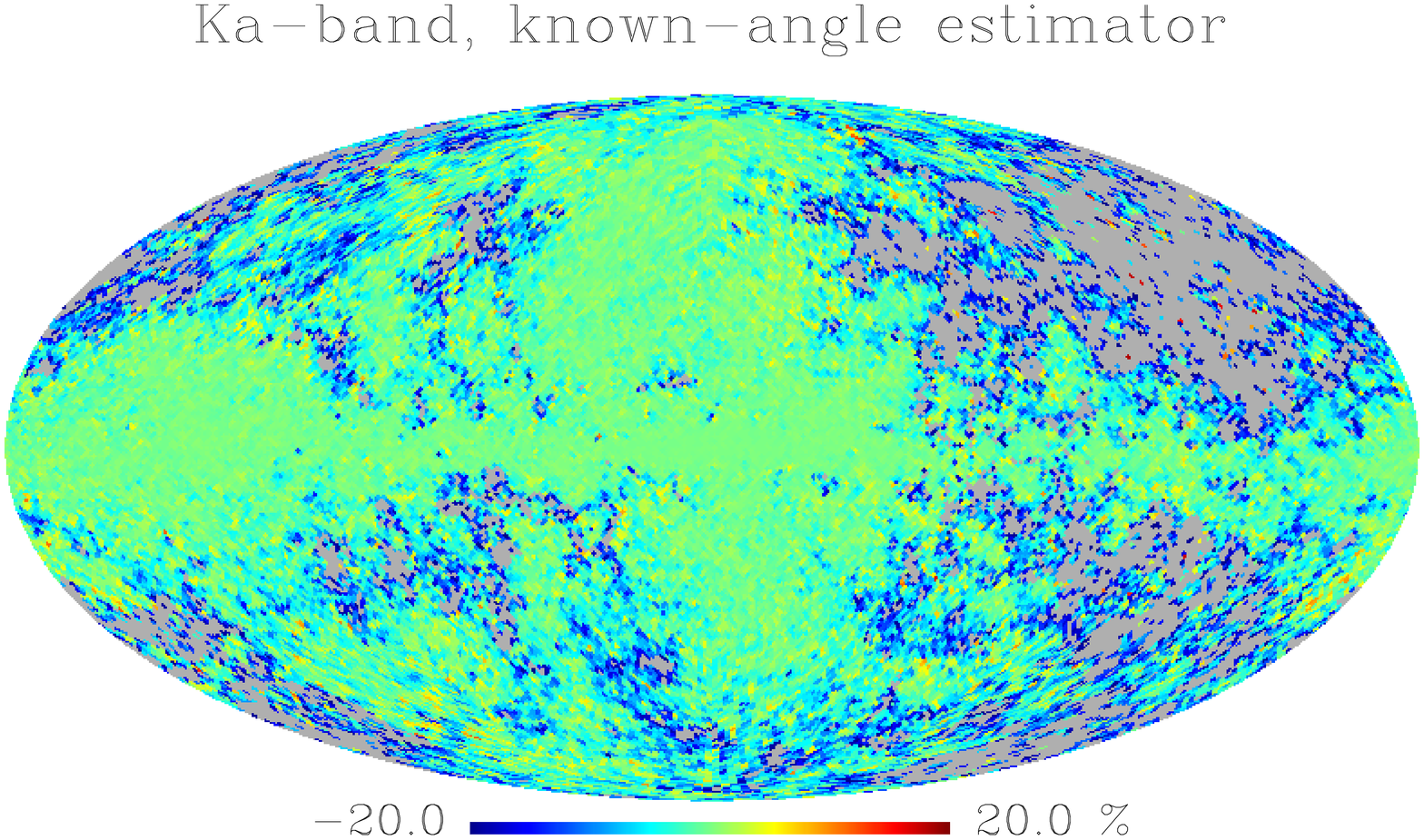}
   \includegraphics[angle=90,width=\widthfig\textwidth]{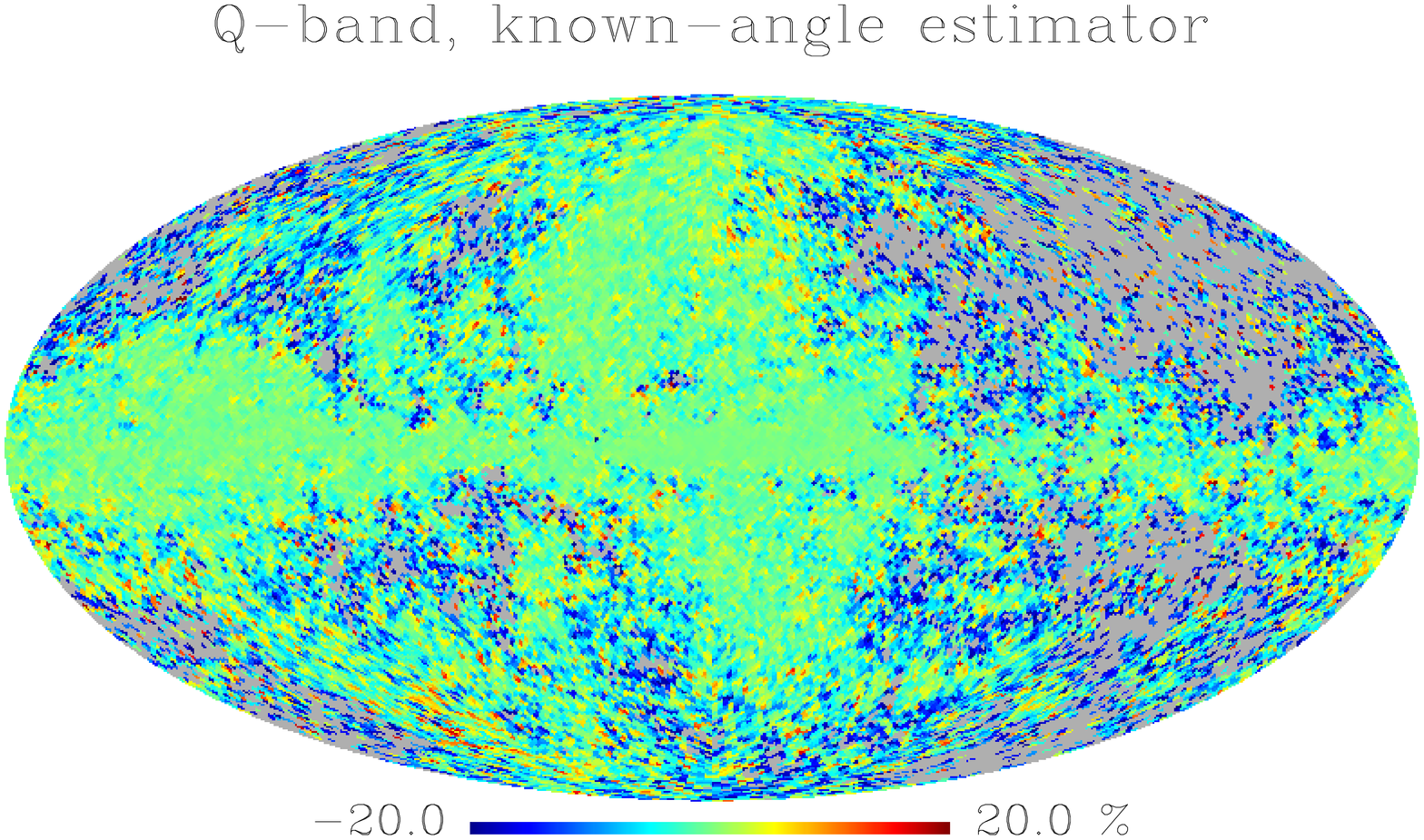}
   \caption{Maps showing the fractional bias in the \wmap K, Ka- and
     Q-bands (left, middle and right columns respectively).  The top
     {\it top} row shows the fractional bias for the naive estimator
     $P'=\sqrt{Q'^2+U'^2}$.  The second row shows the residual bias
     when using the $\pmas$ estimator and the {\it bottom} row has the
     residual bias that remains after correcting with the $\pka$
     estimator. All the pixels with an absolute value larger than 0.2
     are shown in grey. }
   \label{fig:bias_maps}
\end{figure*}

In order to see where the residual bias is more important, we show
maps of the fractional bias after the correction using the different
estimators. Fig.~\ref{fig:bias_maps} shows the fractional bias at K-,
Ka- and Q-band, for each of our estimators. Pixels where the residual
bias is larger than 20 per cent are shown in grey. 
The $\pmas$ estimator (second row) leaves a small residual bias over
most of the sky (green areas in the Figure).  The $\pka$ estimator
(bottom row) performs clearly better. In
Table\,\ref{tab:area_bias0.2} we list the percentage of the area of
the sky with a residual fractional bias smaller than $\pm 0.2$.

\begin{table}
  \caption{Percentage of the area of the full-sky map that have an
    absolute value of the residual fractional bias smaller than
    0.2. These areas corresponds to the coloured pixels in
    Fig. \ref{fig:bias_maps}.}  \centering
  \begin{tabular}{lccc}
    \toprule
    Estimator & \kband & \kaband & \qband \\
    &  \% &  \% & \%  \\
    \midrule
    $ P'$    & 84.9   & 35.0 &  18.4  \\
    $ \pmas$ & 91.5   & 53.3 &  29.5  \\ 
    $ \pka$  & --     & 84.8 &  84.3  \\
    \bottomrule
  \end{tabular}
   \label{tab:area_bias0.2}   
\end{table}

The excellent performance of the known-angle estimator in this case is
due to the use of some extra information. The improvement is
spectacular due to the very low SNR over much of the sky at the target
frequencies.  One could attempt to restore the signal-to-noise by
simply smoothing to lower resolution, but this will fail when there is
substantial real variation in the polarisation angle, which would
cause the $Q$ and $U$ signals averaged over large areas to tend
towards zero. The great advantage of the known-angle estimator is that
we correctly preserve the true polarisation direction while still
allowing coherent averaging of the polarised amplitude.
\citet{Vidal2015} uses $\pka$ to measure the spectral index of the
diffuse synchrotron polarised emission over large regions of the sky,
an ideal application of the known-angle estimator.  It can also be
applied in the \planck data, both for the low-frequency synchrotron
emission and also for the high-frequency dust polarisation, where a
natural template would be the highest available frequency, since the
dust emission rises with frequency \citep{PIP_XXII}.

\section{Conclusions}
\label{sec:conclusions}

We have proposed a new way of correcting for the positive
bias that affects the polarisation amplitude. The `known-angle
estimator', $\pka$, works when there is independent and high SNR
information about the polarisation angle of the observed source. This
additional information helps to reduce the polarisation amplitude
bias.

We have derived formulae for the estimator, its precision, and its
residual bias, due to the uncertainty in the template angle that is
used. The estimator is continuous, analytic, possibly negative and
highly Gaussian. Given an independent good template for the
polarisation angle, i.e. the SNR of the template is at least two time
the SNR of the target, this estimator performs excellently in the low
SNR regime. We have shown with simulations using \wmap data that the known-angle
estimator, $\pka$, outperforms the modified asymptotic (MAS)
estimators, which is not surprise as $\pka$ uses additional
information about the $U/Q$ ratio to correct for the bias.  We believe
that this new estimator will be of great use in datasets that
encompass multiple frequencies with different SNR rations like
\planck\!\!, or even in multi-wavelength analysis mixing optical,
infrared and radio data.

\section*{Acknowledgements}
We thank the referee, S. Plaszczynski for very useful comments that
have greatly improved the quality and presentation of this paper. We
also thank Anthony Banday and Michael Keith for very useful comments
on this work. MV acknowledges the funding from Becas Chile. CD
acknowledges an STFC Advanced Fellowship, an EU Marie-Curie IRG grant
under the FP7 and ERC Starting Grant (no. 307209).  We acknowledge the
use of the Legacy Archive for Microwave Background Data Analysis
(LAMBDA). Support for LAMBDA is provided by the NASA Office of Space
Science. Some of the work of this paper was done using routines from
the IDL Astronomy User's Library. Some of the results in this paper
have been derived using the HEALPix \citep{gorsky:05} package.

\label{lastpage}

\bibliographystyle{mn2e}

\bibliography{refs}

\bsp
\end{document}